\documentclass[article,showpacs]{revtex4}
\usepackage{graphicx,natbib}
\usepackage{amsmath}
\newcommand{\bi}[1]{\mbox{\boldmath ${#1}$}}
\newcommand{\dis}{\displaystyle}
\newcommand{\lra}{\leftrightarrow}
\begin{document}
\title{$\phi\to\pi^0\eta\gamma$ and $\phi\to\pi^0\pi^0\gamma$ decays \\
and \\
Mixing Between Low and High Mass Scalar Mesons }
\vspace*{1cm}
\author{T. Teshima}
\email{teshima@isc.chubu.ac.jp}
\author{I. Kitamura}
\author{N. Morisita}
\affiliation{Department of Natural Sciences,  Chubu University, Kasugai 
487-8501, Japan\\
\\
}
\begin{abstract}
Radiative decays $\phi\to\eta\pi^0\gamma$ and $\phi\to\pi^0\pi^0\gamma$ 
are studied assuming that these decays are caused through the intermediate $a_0(980)
\gamma$ and $f_0(980)\gamma$ states, respectively. 
Fitting the experimental data of the $\eta\pi^0$ and $\pi^0\pi^0$ invariant mass 
spectrum in the decays $\phi\to\eta\pi^0\gamma$ and $\pi^0\pi^0\gamma$, it is shown 
that the processes $\phi\to a_0\gamma$ and $\phi\to f_0\gamma$ are dominated by the 
$K^+K^-$ loop interaction rather than the contact $\phi a_0(f_0)\gamma$ one 
both for the non-derivative and derivative $SPP$ coupling. 
The experimental data of $\Gamma[\phi\to f_0\gamma]/\Gamma[\phi\to a_0\gamma]$ 
predicts that $g_{f_0K\bar{K}}/g_{a_0K\bar{K}}\sim2$.
Considering the effects of the mixing between low mass scalar $qq\bar{q}\bar{q}$ 
states and high mass scalar $q\bar{q}$ states to these coupling constants $g_{f_0K
\bar{K}}$ and $g_{a_0K\bar{K}}$, one suggests that the mixing is rather large. 
\end{abstract}
\pacs{12.39.Mk, 12.40.Yx, 13.66.Jn}
\preprint{CU-TP/07-02}
\maketitle
%
\section{Introduction}
For a long time, the radiative decays of the $\phi$ to $\pi^0\eta\gamma$ and $\pi^0
\pi^0\gamma$ have been analyzed, assuming that the decays $\phi\to\pi^0\eta(\pi^0)
\gamma$ proceed through the $\phi\to a_0(f_0)\gamma$ decays, to reveal the 
structure of the light scalar mesons $a_0(980)$ and $f_0(980)$ \cite{phi decay I, 
phi decay II, phi decay III}. 
These analyses are performed assuming the charged $K^+K^-$ loop diagram in the 
coupling $\phi a_0(f_0)\gamma$ and the contact (vector dominance) $\phi a_0(f_0)\gamma$ 
interaction. 
In this work, we analyze the data for the $\pi^0\eta$ and $\pi^0\pi^0$ invariant 
mass spectrum in $\phi\to \pi^0\eta\gamma$ and $\phi\to \pi^0\pi^0\gamma$ 
decays given in recent precise experiment \cite{rare decay}, assuming charged 
$K^+K^-$ loop diagram and contact $\phi f_0(a_0)\gamma$ interaction. 
For the $S$(scalar meson)-$P$(pseudo-scalar meson)-$P$(pseudo-scalar meson) 
interactions appeared in these decaying processes, we consider the cases 
non-derivative interaction and derivative one, latter of which is adopted 
in the literature \cite{phi decay III}. 
The result obtained in our present analysis shows that these processes are caused 
through the $K^+K^-$ loop diagram dominantly. 
\par
When the $K^+K^-$ loop diagram is dominant in the decays $\phi\to f_0 \gamma\to
\pi^0\pi^0\gamma$ and $\phi\to a_0 \gamma\to\pi^0\eta\gamma$, the value of the 
ratio $g_{f_0K\bar{K}}/g_{a_0K\bar{K}}$ is obtained from the experimental ratio 
$\Gamma(\phi\to a_0\gamma)/\Gamma(\phi\to f_0\gamma)$, 
that is the rather large value $g_{f_0 K\bar{K}}/g_{a_0K\bar{K}}\sim2$. 
These coupling constant strengths depend on the structure of the scalar 
mesons, that is, these scalar mesons are constituted of $q\bar{q}$ or 
$qq\bar{q}\bar{q}$, or are mixing states of $q\bar{q}$ and $qq\bar{q}\bar{q}$. 
Authors in Ref. \cite{phi decay I} argue that the data of $\phi\to 
a_0(f_0)\gamma\to\pi^0\eta(\pi^0)\gamma$ decays gives evidence in favor of the 
$qq\bar{q}\bar{q}$ nature for the scalar $a_0(980)$ and $f_0(980)$ mesons, 
and authors in Ref. \cite{phi decay II} argue the matter of mixing between 
$q\bar{q}$ and $qq\bar{q}\bar{q}$ states. 
\par
Recent many literature (refer to the  "Note on scalar mesons"  in 
\cite{PDG 2006}) suggest that the low mass scalar nonet ($f_0(600)$, $K^*_0(800)$, 
$a_0(980)$, $f_0(980)$) are $qq\bar{q}\bar{q}$ state and the high mass scalar 
mesons ($a_0(1450)$, $K^*_0(1430)$, $f_0(1370)$, $f_0(1500)$, $f_0(1710)$) are 
conventional $L=1$ $q\bar{q}$ nonet plus one glueball. 
We assume a strong mixing between low mass and high mass scalar mesons to explain 
the fact that the high $L=1$ $q\bar{q}$ scalar nonet are so high compared to other 
$L=1\ q\bar{q}\ 1^{++}$ and $2^{++}$ mesons \cite{mixing,teshima1}. 
Assuming that the coupling strengths causing the mixing between $I=1\ a_0(980)$ 
and $a_0(1450)$, $I=1/2\ K^*_0(800)$ and $K^*_0(1430)$ and $I=0$ ($f_0(600)$, 
$f_0(980)$) and ($f_0(1370)$, $f_0(1500)$, $f_0(1710)$) are same, we analyzed the 
$S\to PP$ decays using  derivative $SPP$ couplings \cite{teshima2}. 
Fitting the various experimental $SPP$ decay widths, we obtained the mixing angle 
between $a_0(980)$ and $a_0(1450)$ as ${\sim} 9^{\circ}$. 
In our previous work \cite{teshima3}, we analyzed the $\Gamma(\phi\to 
f_0\gamma)$ and $\Gamma(\phi\to a_0\gamma)$ assuming the contact (vector dominance)
coupling for the $a_0\phi\gamma$ and $f_0\phi\gamma$ interaction and using the 
mixing strength obtained in previous work \cite{teshima2}, and then suggested the 
importance of the mixing effect for the explanation of the rather large ratio 
$\Gamma(\phi\to f_0\gamma)/\Gamma(\phi\to a_0\gamma)$. 
\par
In section 2, we analyze the data for $\pi^0\eta$ and $\pi^0\pi^0$ invariant mass 
spectrum of the $d\,BR(\phi\to \pi^0\pi^0\gamma)/dq$ and $d\,BR(\phi\to 
\pi^0\eta\gamma)/dq$ assuming the intermediate scalar states 
$f_0(980)$ and $a_0(980)$. 
In this analyses, we consider the contact and $K^+K^-$ loop interaction for 
$\phi f_0(a_0)\gamma$ coupling, in cases of derivative and non-derivative $SPP$ 
coupling. 
In section 3, we reanalyze our mass formula for low mass nonet scalar and high 
mass nonet scalar + glueball adopting the new mass data of $K^*_0(800)$ 
\cite{PDG 2006}. 
In section 4, we express the $SPP$ coupling constants $g_{a_0\pi\pi}$, $g_{f_0\pi\pi}$ 
etc. using the mixing parameters between low and high mass scalar mesons. 
We pursue the best fit analyses for the $S\to PP$ decay data using the mixing 
parameters obtained in section 3 for both non-derivative and derivative $SPP$ 
interactions, and then obtain the best fit $g_{f_0K\bar{K}}$ etc. 
Comparing the best fit $g_{f_0K\bar{K}}$ etc. with the values obtained from 
the $\phi\to f_0(a_0)\gamma$ decays, we suggest that the non-derivative coupling is 
reasonable than the derivative one and the mixing between $q\bar{q}$ and 
$qq\bar{q}\bar{q}$ is rather large.
%
%
\section{Analysis of the $\phi\to\pi^0\eta\gamma$ and $\phi\to\pi^0\pi^0\gamma$ 
decays}
\subsection{$\phi\to\pi^0\eta\gamma$ decay} 
Firstly, we consider the decay $\phi\to a_0(980)\gamma\to\pi^0\eta\gamma$ shown 
in Fig. 1. 
\begin{figure}
\vspace{0.5cm}
\begin{center}
\unitlength 0.1in
\begin{picture}( 26.0000, 20.0000)(  4.0000,-26.0000)
%
\special{pn 8}%
\special{ar 1400 1600 100 100  0.0000000 6.2831853}%
%
\special{pn 13}%
\special{pa 400 1600}%
\special{pa 1300 1600}%
\special{fp}%
\special{pa 1480 1650}%
\special{pa 2200 2000}%
\special{fp}%
\special{pa 2200 2000}%
\special{pa 3000 2000}%
\special{fp}%
\special{pa 2200 2000}%
\special{pa 2800 2600}%
\special{fp}%
\special{pn 13}%
\special{pa 1484 1544}%
\special{pa 1492 1548}%
\special{pa 1500 1552}%
\special{pa 1508 1556}%
\special{pa 1516 1560}%
\special{pa 1524 1564}%
\special{pa 1532 1566}%
\special{pa 1540 1568}%
\special{pa 1546 1568}%
\special{pa 1552 1568}%
\special{pa 1556 1568}%
\special{pa 1562 1566}%
\special{pa 1566 1562}%
\special{pa 1570 1558}%
\special{pa 1572 1552}%
\special{pa 1574 1546}%
\special{pa 1576 1540}%
\special{pa 1578 1532}%
\special{pa 1580 1524}%
\special{pa 1580 1514}%
\special{pa 1582 1506}%
\special{pa 1582 1496}%
\special{pa 1582 1486}%
\special{pa 1584 1478}%
\special{pa 1584 1468}%
\special{pa 1586 1460}%
\special{pa 1586 1452}%
\special{pa 1588 1446}%
\special{pa 1592 1440}%
\special{pa 1594 1434}%
\special{pa 1598 1430}%
\special{pa 1602 1426}%
\special{pa 1606 1424}%
\special{pa 1612 1424}%
\special{pa 1618 1424}%
\special{pa 1624 1424}%
\special{pa 1632 1426}%
\special{pa 1638 1428}%
\special{pa 1646 1432}%
\special{pa 1654 1434}%
\special{pa 1662 1438}%
\special{pa 1672 1442}%
\special{pa 1680 1446}%
\special{pa 1688 1450}%
\special{pa 1696 1454}%
\special{pa 1704 1458}%
\special{pa 1712 1460}%
\special{pa 1718 1462}%
\special{pa 1726 1464}%
\special{pa 1732 1464}%
\special{pa 1736 1462}%
\special{pa 1742 1462}%
\special{pa 1746 1458}%
\special{pa 1750 1454}%
\special{pa 1754 1450}%
\special{pa 1756 1444}%
\special{pa 1758 1436}%
\special{pa 1760 1430}%
\special{pa 1760 1422}%
\special{pa 1762 1412}%
\special{pa 1762 1404}%
\special{pa 1764 1394}%
\special{pa 1764 1384}%
\special{pa 1764 1376}%
\special{pa 1766 1366}%
\special{pa 1766 1358}%
\special{pa 1768 1350}%
\special{pa 1770 1342}%
\special{pa 1772 1336}%
\special{pa 1774 1330}%
\special{pa 1778 1326}%
\special{pa 1782 1322}%
\special{pa 1786 1320}%
\special{pa 1792 1318}%
\special{pa 1798 1318}%
\special{pa 1804 1318}%
\special{pa 1810 1320}%
\special{pa 1818 1322}%
\special{pa 1826 1326}%
\special{pa 1834 1328}%
\special{pa 1842 1332}%
\special{pa 1850 1336}%
\special{pa 1858 1340}%
\special{pa 1866 1346}%
\special{pa 1876 1348}%
\special{pa 1884 1352}%
\special{pa 1890 1356}%
\special{pa 1898 1358}%
\special{pa 1904 1358}%
\special{pa 1912 1360}%
\special{pa 1916 1358}%
\special{pa 1922 1358}%
\special{pa 1926 1354}%
\special{pa 1930 1350}%
\special{pa 1934 1346}%
\special{pa 1936 1340}%
\special{pa 1938 1334}%
\special{pa 1940 1326}%
\special{pa 1942 1318}%
\special{pa 1942 1310}%
\special{pa 1944 1302}%
\special{pa 1944 1292}%
\special{pa 1944 1282}%
\special{pa 1946 1274}%
\special{pa 1946 1264}%
\special{pa 1948 1256}%
\special{pa 1948 1248}%
\special{pa 1950 1240}%
\special{pa 1952 1234}%
\special{pa 1956 1228}%
\special{pa 1958 1222}%
\special{pa 1962 1218}%
\special{pa 1966 1216}%
\special{pa 1972 1214}%
\special{pa 1978 1214}%
\special{pa 1984 1214}%
\special{pa 1990 1216}%
\special{pa 1998 1218}%
\special{pa 2004 1220}%
\special{pa 2012 1224}%
\special{pa 2020 1226}%
\special{pa 2030 1230}%
\special{pa 2038 1236}%
\special{pa 2046 1240}%
\special{pa 2054 1244}%
\special{pa 2062 1246}%
\special{pa 2070 1250}%
\special{pa 2078 1252}%
\special{pa 2084 1254}%
\special{pa 2090 1254}%
\special{pa 2096 1254}%
\special{pa 2102 1252}%
\special{pa 2106 1250}%
\special{pa 2110 1248}%
\special{pa 2114 1242}%
\special{pa 2118 1238}%
\special{pa 2120 1232}%
\special{pa 2122 1224}%
\special{pa 2122 1216}%
\special{pa 2124 1208}%
\special{pa 2124 1198}%
\special{pa 2126 1190}%
\special{pa 2126 1180}%
\special{pa 2126 1172}%
\special{pa 2128 1162}%
\special{pa 2128 1154}%
\special{pa 2130 1144}%
\special{pa 2132 1138}%
\special{pa 2134 1130}%
\special{pa 2136 1124}%
\special{pa 2138 1120}%
\special{pa 2142 1116}%
\special{pa 2146 1112}%
\special{pa 2152 1110}%
\special{pa 2156 1108}%
\special{pa 2162 1110}%
\special{pa 2170 1110}%
\special{pa 2176 1112}%
\special{pa 2184 1114}%
\special{pa 2192 1118}%
\special{pa 2200 1122}%
\special{pa 2208 1126}%
\special{pa 2216 1130}%
\special{pa 2226 1134}%
\special{pa 2234 1138}%
\special{pa 2242 1140}%
\special{pa 2250 1144}%
\special{pa 2256 1146}%
\special{pa 2264 1148}%
\special{pa 2270 1150}%
\special{pa 2276 1150}%
\special{pa 2282 1148}%
\special{pa 2286 1146}%
\special{pa 2290 1144}%
\special{pa 2294 1140}%
\special{pa 2298 1134}%
\special{pa 2300 1128}%
\special{pa 2302 1122}%
\special{pa 2304 1114}%
\special{pa 2306 1106}%
\special{pa 2306 1096}%
\special{pa 2306 1088}%
\special{pa 2308 1078}%
\special{pa 2308 1068}%
\special{pa 2308 1060}%
\special{pa 2310 1050}%
\special{pa 2310 1042}%
\special{pa 2312 1034}%
\special{pa 2314 1028}%
\special{pa 2316 1022}%
\special{pa 2320 1016}%
\special{pa 2322 1012}%
\special{pa 2326 1008}%
\special{pa 2332 1006}%
\special{pa 2336 1004}%
\special{pa 2342 1004}%
\special{pa 2350 1004}%
\special{pa 2356 1006}%
\special{pa 2364 1008}%
\special{pa 2372 1012}%
\special{pa 2380 1016}%
\special{pa 2388 1020}%
\special{pa 2396 1024}%
\special{pa 2404 1028}%
\special{pa 2412 1032}%
\special{pa 2420 1036}%
\special{pa 2428 1038}%
\special{pa 2436 1042}%
\special{pa 2444 1044}%
\special{pa 2450 1044}%
\special{pa 2456 1044}%
\special{pa 2462 1044}%
\special{pa 2466 1042}%
\special{pa 2472 1040}%
\special{pa 2476 1036}%
\special{pa 2478 1032}%
\special{pa 2480 1026}%
\special{pa 2484 1018}%
\special{pa 2484 1012}%
\special{pa 2486 1004}%
\special{pa 2488 994}%
\special{pa 2488 986}%
\special{pa 2488 976}%
\special{pa 2490 966}%
\special{pa 2490 958}%
\special{pa 2490 948}%
\special{pa 2492 940}%
\special{pa 2494 932}%
\special{pa 2494 924}%
\special{pa 2498 918}%
\special{pa 2500 912}%
\special{pa 2504 908}%
\special{pa 2506 904}%
\special{pa 2512 902}%
\special{pa 2516 900}%
\special{pa 2522 900}%
\special{pa 2528 900}%
\special{pa 2536 902}%
\special{pa 2542 904}%
\special{pa 2550 906}%
\special{pa 2558 910}%
\special{pa 2566 914}%
\special{pa 2574 918}%
\special{pa 2584 922}%
\special{pa 2592 926}%
\special{pa 2600 930}%
\special{pa 2608 934}%
\special{pa 2616 936}%
\special{pa 2622 938}%
\special{pa 2630 940}%
\special{pa 2636 940}%
\special{sp}%
%
\special{pn 8}%
\special{pa 1470 1530}%
\special{pa 1330 1670}%
\special{fp}%
\special{pa 1490 1570}%
\special{pa 1370 1690}%
\special{fp}%
\special{pa 1490 1630}%
\special{pa 1430 1690}%
\special{fp}%
\special{pa 1430 1510}%
\special{pa 1310 1630}%
\special{fp}%
\special{pa 1370 1510}%
\special{pa 1310 1570}%
\special{fp}%
%
\special{pn 13}%
\special{pa 800 1600}%
\special{pa 850 1600}%
\special{fp}%
\special{sh 1}%
\special{pa 850 1600}%
\special{pa 784 1580}%
\special{pa 798 1600}%
\special{pa 784 1620}%
\special{pa 850 1600}%
\special{fp}%
%
\special{pn 13}%
\special{pa 2500 2000}%
\special{pa 2700 2000}%
\special{fp}%
\special{sh 1}%
\special{pa 2700 2000}%
\special{pa 2634 1980}%
\special{pa 2648 2000}%
\special{pa 2634 2020}%
\special{pa 2700 2000}%
\special{fp}%
%
\special{pn 13}%
\special{pa 2200 2000}%
\special{pa 2620 2420}%
\special{fp}%
\special{sh 1}%
\special{pa 2620 2420}%
\special{pa 2588 2360}%
\special{pa 2582 2382}%
\special{pa 2560 2388}%
\special{pa 2620 2420}%
\special{fp}%
\put(6.7000,-15.3000){\makebox(0,0)[lb]{$\phi(p,\epsilon_\phi)$}}%
\put(15.9000,-19.9000){\makebox(0,0)[lb]{$a_0 (q)$}}%
\put(29.2000,-19.5000){\makebox(0,0)[lb]{$\pi^0(q_1)$}}%
\put(28.7000,-26.6000){\makebox(0,0)[lb]{$\eta(q_2)$}}%
\put(27.0000,-9.5000){\makebox(0,0)[lb]{$\gamma(k,\epsilon_\gamma)$}}%
%
\special{pn 20}%
\special{sh 1}%
\special{ar 2200 2000 10 10 0  6.28318530717959E+0000}%
\special{sh 1}%
\special{ar 2200 2000 10 10 0  6.28318530717959E+0000}%
\special{sh 1}%
\special{ar 2200 2000 10 10 0  6.28318530717959E+0000}%
\end{picture}%
\hspace{2cm} \\
\caption{
Diagram for the decay $\phi\to a_0(980)\gamma\to\pi^0\eta\gamma$
}
\end{center}
\end{figure}
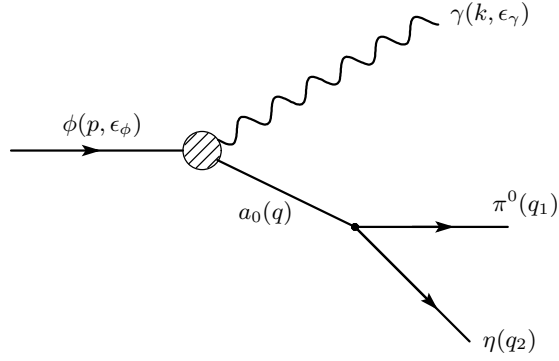
The invariant mass distribution of the branching ratio $d\,BR(\phi\to a_0(980)
\gamma\to\pi^0\eta\gamma)/dm$ is expressed as (refer to first paper in 
Ref. \cite{phi decay I})
\begin{equation}
\frac{d\,BR(\phi\to a_0\gamma\to\pi^0\eta\gamma)}{dm}=
\frac{2m^2}{\pi}\frac{1}{\Gamma_{\phi}}\frac{\Gamma(\phi\to a_0\gamma:m)
\Gamma(a_0\to\pi^0\eta:m)}{|D_{a_0}(m^2)|^2}, 
\end{equation}
where $\Gamma_\phi$ is a decay width of $\phi$ and $1/D_{a_0}(m^2)$ represents 
the propagator of the intermediate state $a_0$,
\begin{equation}
D_{a_0}(m^2)=m^2-m_{a_0}^2-im_{a_o}\Gamma_{a_0}.
\end{equation} 
$\Gamma(a_0\to\pi^0\eta:m)$ is the decay width on the virtual mass $m$ of 
intermediate $a_0$ defined as 
$m=\sqrt{q_0^2-{\bf q}^2}$, 
\begin{eqnarray}
 \Gamma(a_0\to\pi^0\eta:m)&=&
 \dis{\frac{g_{a_0\pi\eta}^2}{8\pi m^2}\frac{\sqrt{(m^2-(m_{\pi}+m_{\eta})^2)
 (m^2-(m_{\pi}-m_{\eta})^2)}}{2m}} \nonumber \\ 
 &&\times\left\{\begin{array}{l}
 1\ \ \ \ {\rm for\ non{\mbox{-}}derivative\ coupling,}\\
 \dis{\left(\frac{m^2-m_{\pi}^2-m_{\eta}^2}{2}\right)^2}\ \ \ \ {\rm for\ 
 derivative\ coupling,}
 \end{array}\right.
\end{eqnarray} 
where coupling constant $g_{a_{0}\pi\eta}$ is defined as 
\begin{equation}
M(a_0(q)\to\pi^0(q_1)+\eta(q_2))=g_{a_0\pi\eta}\times\left\{
\begin{array}{l}
1\ \ \ \ {\rm for\ non{\mbox{-}}derivative\ coupling,}\\
q_1\cdot q_2\ \ \ \ {\rm for\ 
 derivative\ coupling.}
\end{array}\right.
\end{equation}
$\Gamma(\phi\to a_0\gamma:m)$ is the decay width on the virtual mass $m=
\sqrt{q_0^2-{\bf q}^2}$ of intermediate state $a_0$, 
\begin{equation}
 \dis{\Gamma(\phi\to a_0\gamma:m)=\frac{\alpha}{3}
 g_{\phi a_0\gamma}^2(m)\left(\frac{m_{\phi}^2-m^2}{2m_{\phi}}
 \right)^3},
\end{equation} 
where coupling constant $g_{\phi a_0\gamma}(m)$ is defined as 
\begin{equation}
M(\phi(p,\epsilon_\phi)\to a_0(q)+\gamma(k,\epsilon_\gamma))=
eg_{\phi a_0\gamma}(m)(p\cdot k\epsilon_\phi\cdot\epsilon_\gamma-
p\cdot\epsilon_\gamma k\cdot\epsilon_\phi).
\end{equation} 
\par
For the coupling $g_{\phi a_0\gamma}(m)$, contact interaction and $K^+K^-$ loop 
interaction contribute as shown in Fig. 2, and then $g_{\phi a_0\gamma}(m)$ is 
expressed as 
\begin{equation}
g_{\phi a_0\gamma}(m)=g_{\phi a_0\gamma}^{\rm contact}+
g_{\phi a_0\gamma}^{K\bar{K}\,\rm loop}(m).
\end{equation}
\begin{figure}
\vspace{0.5cm}
\begin{center}
\input{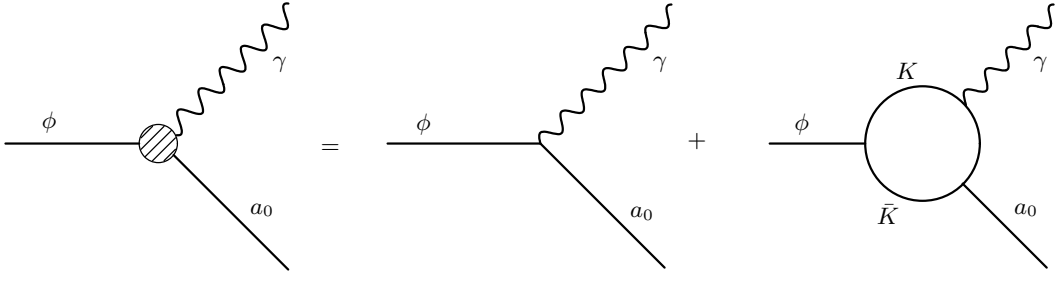}\hspace{2cm} \\
\caption{
Diagrams for the contact and $K^+K^-$ loop coupling contributing to
$g_{\phi a_0\gamma}(m)$
}
\end{center}
\end{figure}
$g_{\phi a_0\gamma}^{K\bar{K}\,\rm loop}(m)$ is calculated for non-derivative 
$a_0K^+K^-$ coupling by many authors (N. N. Achasov {\it et al.} 
and F. E. Close {\it et al.} in \cite{phi decay I}) considering three diagrams 
(a), (b) and (c) shown in Fig. 3, as
\begin{figure}
\vspace{0.5cm}
\begin{center}
\input{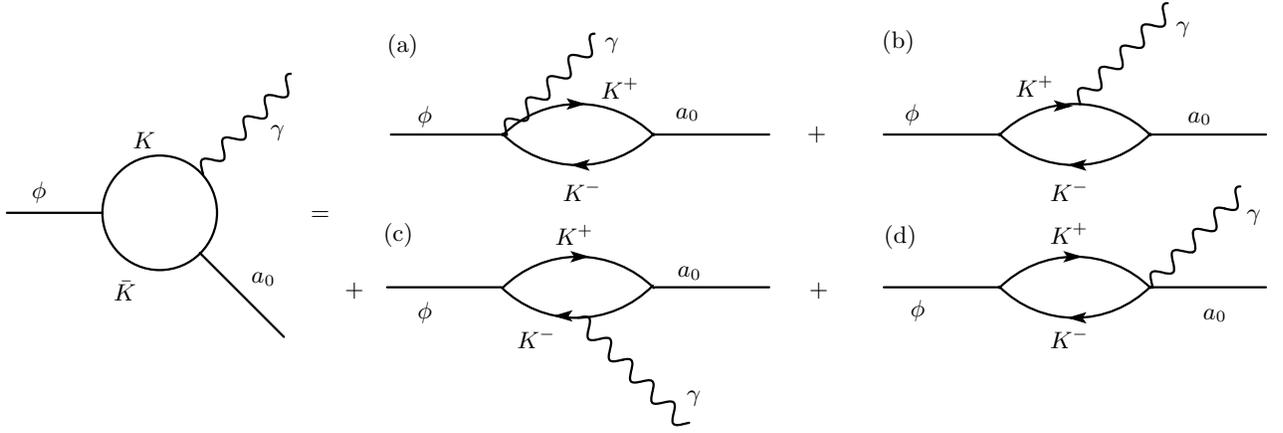}\hspace{2cm} \\
\caption{Diagrams for the $K^+K^-$ loop coupling contributing to $g_{\phi 
a_0\gamma}(m)$ }
\end{center}
\end{figure} 
\begin{equation}
g_{\phi a_0\gamma}^{K\bar{K}\,\rm loop}(m)=\frac{g_{\phi K\bar{K}}g_{a_0K\bar{K}}}
{2\pi^2im_K^2}I(a, b).\ \ \ \ \ {\rm for\ non{\mbox -}derivative\ coupling}
\end{equation} 
The quantities $a$, $b$ are defined as $a=m_\phi^2/m_K^2$, $b=m^2/m_K^2$ 
and $I(a,b)$ arisen from the loop integral is 
\begin{equation}
I(a,b)=\frac{1}{2(a-b)}-\frac{2}{(a-b)^2}\left\{f\left(\frac{1}{b}\right)-
f\left(\frac{1}{a}\right)\right\}+\frac{a}{(a-b)^2}\left\{g\left(\frac{1}{b}
\right)-g\left(\frac{1}{a}\right)\right\},
\end{equation}
where
\begin{eqnarray}
&&f(x)=\left\{\begin{array}{l}
\dis{-\left(\sin^{-1}\left(\frac{1}{2\sqrt{x}}\right)\right)^2},\ \ \ x>\frac14\\
\dis{\frac14\left[\log\frac{\eta_+}{\eta_-}-i\pi\right]^2},\ \ \ x<\frac14
\end{array}\right.\nonumber\\
&&g(x)=\left\{\begin{array}{l}
\dis{\sqrt{4x-1}\sin^{-1}\left(\frac{1}{2\sqrt{x}}\right)},\ \ \ x>\frac14\\
\dis{\frac12\sqrt{1-4x}\left[\log\frac{\eta_+}{\eta_-}-i\pi\right]},\ \ \ x<\frac14 
\end{array}\right.\nonumber\\
&&\ \ \ \ \eta_{\pm}=\frac{1}{2x}\left(1\pm\sqrt{1-4x}\right).
\end{eqnarray}
The coupling constant $g_{\phi K\bar{K}}$ is defined as 
\begin{equation}
M(\phi(p, \epsilon^\phi)\to K^+(q_1)+K^-(q_2))=g_{\phi K\bar{K}}
\epsilon^\phi_\mu(q_1^\mu-q_2^\mu),
\end{equation}
and decay width is expressed as 
\begin{equation}
\Gamma(\phi\to K^++K^-)=\frac{g_{\phi K\bar{K}}^2}{4\pi}\frac{2}{3m_\phi^2}
\left(\frac{m_\phi^2}{4}-m_K^2\right)^{3/2}.
\end{equation}
The coupling constant $g_{\phi K\bar{K}}$ is estimated using the experimental 
data $\Gamma(\phi\to K^++K^-)=2.10\pm0.05 {\rm MeV}$ \cite{PDG 2006} as 
\begin{equation}
g_{\phi K\bar{K}}=4.55\pm0.06.
\end{equation}
For the $a_0K^+K^-$ coupling, $g_{a_0K\bar{K}}$ is defined by the 
similar expression as Eq. (4)
$$
\hspace{2.5cm}M(a_0(q)\to K^+(q_1)+K^-(q_2))=g_{a_0K\bar{K}}\times\left\{
\begin{array}{l}
1\ \ \ \ {\rm for\ non{\mbox{-}}derivative\ coupling,}\\
q_1\cdot q_2\ \ \ \ {\rm for\ 
 derivative\ coupling.}
\end{array}\right.\hspace{2.5cm} (4')
$$
For derivative coupling of $a_0K^+K^-$, $K^+K^-$ loop diagram contribution 
$g_{\phi a_0\gamma}^{K\bar{K}\,\rm loop}(m)$ is calculated by D. Black 
{\it et al.} \cite{phi decay II} considering four diagrams (a), (b), (c) and (d) 
shown in Fig. 3, as 
$$
\hspace{3cm} g_{\phi a_0\gamma}^{K\bar{K}\,\rm loop}(m)=\frac{g_{\phi 
K\bar{K}}g_{a_0K\bar{K}}}{2\pi^2im_K^2}\frac{2m_K^2-m^2}{2}I(a, b)\ \ \ \ \ 
{\rm for\ derivative\ coupling.}\hspace{3cm} (8')
$$ 
\par
Using Eqs. (2), (3), (5), (7), (8), (8'), we parameterize Eq. (1) 
as 
\begin{eqnarray}
&&\frac{d\,BR(\phi\to a_0\gamma\to\pi^0\eta\gamma)}{dm}=G_1
\frac{|G_2+\frac1i\left[\frac{2m_K^2-m^2}{2}\right]I(a,b)|^2}
{|G_2+\frac1i\left[\frac{2m_K^2-m_a^2}{2}\right]I(a,b_0)|^2}
\left(\frac{m_\phi^2-m^2}{m_\phi^2-m_a^2}\right)^3\frac{m_a}{m} \nonumber\\ 
&&\ \ \ \ \ \ \ \ \ \ \ \ \ \ \ \ \times\frac{m_a^2\Gamma_a^2}{(m^2-m_a^2)^2+m_a^2
\Gamma_a^2}\sqrt{\frac{(m^2-(m_\eta+m_\pi)^2)(m^2-(m_\eta-m_\pi)^2)}
{(m_a^2-(m_\eta+m_\pi)^2)(m_a^2-(m_\eta-m_\pi)^2)}}, 
\end{eqnarray}
where $G_1, G_2, b_0$ are defined as 
\begin{equation}
\begin{array}{l}
G_1=\dis{\frac{2}{\pi\Gamma_\phi\Gamma_a^2}}\Gamma(\phi\to a_0\gamma:m_a)
\Gamma(a_0\to \eta\pi^0:m_a),\\
G_2=\dis{g^{\rm contact}_{\phi\gamma a}/\left(\frac{g_{\phi K\bar{K}}g_{a_0K\bar{K}}}
{2\pi^2m_K^2}\right)},\\
b_0=\dis{\frac{m_a^2}{m_K^2}},
\end{array}
\end{equation}
and factors $\left[\frac{2m_K^2-m^2}{2}\right]$ and $\left[\frac{2m_K^2-m_a^2}{2}
\right]$ are replaced to 1 for non-derivative $SPP$ coupling. 
$\Gamma(a_0\to \eta\pi^0, m_a)$ and $\Gamma(\phi\to a_0\gamma,m_a)$ are defined in 
Eq. (3) and (5) settling $m\to m_a$. 
We fit the Eq. (14) varying the parameters $G_1$ and $G_2$ using the experimental 
data from the SND collaboration and KLEO collaboration in Ref. \cite{rare decay}. 
Best-fitted curves are shown in Fig. 4; solid line for non-derivative $SPP$ 
coupling and dashed line for derivative $SPP$ coupling are obtained for the choice 
of the parameters $G_1$ and $G_2$ as $G_1=4.1\times10^{-4}{\rm GeV}^{-1},\ G_2=
-0.16$ for non-derivative coupling and $G_1=3.9\times10^{-4}{\rm GeV}^{-1},\ 
G_2=0.08$ for derivative coupling. 
For these choices, the estimated $BR(\phi\to\pi^0\eta\gamma)$ are estimated as 
\begin{eqnarray}
\hspace{2cm}&G_1=4.1\times10^{-4}{\rm GeV}^{-1},\ G_2=-0.16,\ &BR(\phi\to\pi^0
\eta\gamma)=7.03\times10^{-5}\nonumber\\ 
&&  \ {\rm for\ non\mbox{-}derivative\ coupling,}\\
&G_1=3.9\times10^{-4}{\rm GeV}^{-1},\ G_2=0.08,\ &BR(\phi\to\pi^0\eta\gamma)
=7.12\times10^{-5}\nonumber\\ 
&&  \ {\rm for\ derivative\ coupling.}\hspace{5.3cm}(16')\nonumber
\end{eqnarray}
\begin{figure}
\begin{center}
\includegraphics[width=12.5cm]{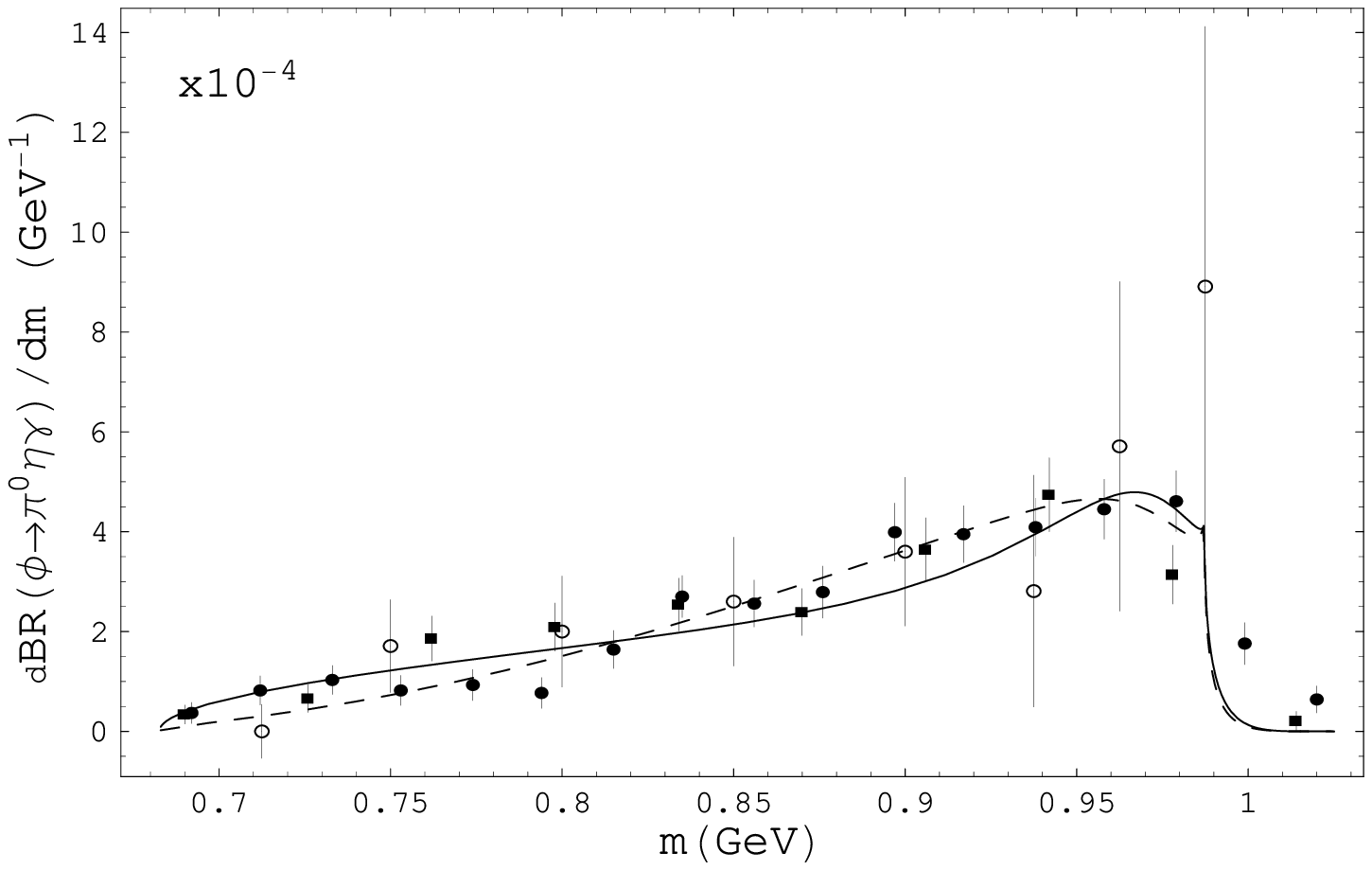}\\
\caption{$dBR(\phi\to\pi^0\eta\gamma)/dm$ (in unit of GeV$^{-1}$) as a function of 
the $\pi^0$-$\eta$ invariant mass $m$ (in GeV$^{-1}$). Solid line shows the best 
fitted curve for the non-derivative $SPP$ coupling interaction and the dashed  
line shows the best fitted curve for the derivative one. 
Experimental data indicated by circles are from the SND collaboration in Ref. 
\cite{rare decay}, and those by filled circles and filled squares are from 
KLEO collaboration in Ref. \cite{rare decay}.
}
\end{center}
\end{figure}
\par
Estimated value for $BR(\phi\to\pi^0\eta\gamma)$ is consistent with the 
experimental data $BR^{\rm exp}(\phi\to\pi^0\eta\gamma)=(8.3\pm0.5)\times10^{-5}$ 
\cite{rare decay, PDG 2006}. Also, estimated value of $G_1$ is consistent with the 
value evaluated from the experimental data (Ref. \cite{PDG 2006}) using the 
relation Eq. (15), 
$$
G_1^{\rm exp}=\dis{\frac{2}{\pi\Gamma_a}}BR^{\rm exp}
(\phi\to a_0\gamma:m_a)BR^{\rm exp}(a_0\to \eta\pi^0:m_a)=(5.96\pm2.47)
\times10^{-4}{\rm GeV}^{-1}.
$$
Furthermore $G_2$ is very small compared to $|I(a, b_0)|=0.902$ for $m_a=0.985$
GeV, then we can suppose that the $K^+K^-$ loop 
contribution is dominant in the $\phi\to\pi^0\eta\gamma$ decay. 
Supposing that the decay $\phi\to a_0\gamma$ is caused through only the $K^+K^-$ 
loop interaction, we obtain the result,
\begin{equation}
 \dis{\Gamma(\phi\to a_0\gamma)=\frac{\alpha}{3}
 \left|\frac{g_{\phi K\bar{K}}g_{a_0K\bar{K}}}{2\pi^2m_K^2}
 \left[\frac{2m_K^2-m_a^2}{2}\right]I(a, b_0)\right|^2\left(\frac{m_{\phi}^2-m_a^2}
 {2m_{\phi}} \right)^3},
\end{equation} 
where the factor $\dis{\left[\frac{2m_K^2-m_a^2}{2}\right]}$ is replaced to 1 for 
the non-derivative coupling. 
Using the value Eq. (13) of $g_{\phi K\bar{K}}$ and the experimental value  
$\Gamma(\phi\to a_0\gamma)=(0.323\pm0.029)\times10^{-3}{\rm MeV}$ in Ref.\cite{
PDG 2006}, we obtain the result
\begin{equation}
g_{a_0K\bar{K}}=\left\{
\begin{array}{l}
2.18\pm0.12\ \ {\rm GeV},\ \ \ \ {\rm for\ non\mbox{-}derivative\ coupling}\\
9.04\pm0.50\ \ {\rm GeV^{-1}}.\ \ \ \ {\rm for \ derivative\ coupling}
\end{array} 
\right.
\end{equation}
Using relations Eqs. (15), (3), (17) and estimated results Eq. (16), (16'), (18), 
we obtained the values for $g_{a_0\pi\eta}$,
$$
\hspace{4cm}g_{a_0\pi\eta}=\left\{
\begin{array}{l}
1.89\pm0.75\ \ {\rm GeV},\ \ \ \ {\rm for\ non\mbox{-}derivative\ coupling}\\
5.79\pm2.32\ \ {\rm GeV^{-1}}.\ \ \ \ {\rm for \ derivative\ coupling}
\end{array} 
\right.\hspace{4cm}(18')
$$
%
\subsection{$\phi\to\pi^0\pi^0\gamma$ decay} 
For the decay $\phi\to f_0\gamma\to\pi^0\pi^0\gamma$, the invariant mass 
distribution of the branching ratio 
$dBR(\phi\to f_0\gamma\to\pi^0\pi^0\gamma)/dm$ is expressed similar to Eq. (14) 
for the case $\phi\to a_0\gamma\to\pi^0\eta\gamma$ as
\begin{eqnarray}
&&\frac{d\,BR(\phi\to f_0\gamma\to\pi^0\pi^0\gamma)}{dm}=G_1
\frac{|G_2+\frac1i\left[\frac{2m_K^2-m^2}{2}\right]I(a,b)|^2}
{|G_2+\frac1i\left[\frac{2m_K^2-m_f^2}{2}\right]I(a,b_0)|^2}
\left(\frac{m_\phi^2-m^2}{m_\phi^2-m_f^2}\right)^3\frac{m_f}{m} \nonumber\\ 
&&\ \ \ \ \ \ \ \ \ \ \ \ \ \ \ \ \times\frac{m_f^2\Gamma_f^2}{(m^2-m_f^2)^2+m_f^2
\Gamma_f^2}\sqrt{\frac{m^2-4m_\pi^2}{m_f^2-4m_\pi^2}}, 
\end{eqnarray}
where $G_1, G_2, b_0$ are defined as 
\begin{equation}
\begin{array}{l}
G_1=\dis{\frac{2}{\pi\Gamma_\phi\Gamma_f^2}}\Gamma(\phi\to f_0\gamma:m_f)
\Gamma(f_0\to \pi^0\pi^0: m_f),\\
G_2=\dis{g^{\rm contact}_{\phi\gamma f}/\left(\frac{g_{\phi K\bar{K}}g_{f_0K\bar{K}}}
{2\pi^2m_K^2}\right)},\\
b_0=\dis{\frac{m_f^2}{m_K^2}}.
\end{array}
\end{equation}
Here, $\Gamma(f_0\to\pi^0\pi^0:m_f)$ and $\Gamma(\phi\to f_0\gamma:m_f)$ are 
expressed as 
\begin{eqnarray}
 \Gamma(f_0\to\pi^0\pi^0:m_f)&=&
 \dis{\frac{g_{f_0\pi\pi}^2}{16\pi m_f^2}\frac{\sqrt{m_f^2-4m_{\pi}^2}}{2} }
 \nonumber \\ 
 &&\times\left\{\begin{array}{l}
 1\hspace{2cm} {\rm for\ non{\mbox{-}}derivative\ coupling,}\\
 \dis{\left(\frac{m_f^2-2m_{\pi}^2}{2}\right)^2}\ \ \ \ {\rm for\ 
 derivative\ coupling,}
 \end{array}\right.
\end{eqnarray} 
where coupling constant $g_{f_{0}\pi\pi}$ is defined as 
\begin{equation}
M(f_0(q)\to\pi^0(q_1)+\pi^0(q_2))=\frac12g_{f_0\pi\pi}\times\left\{
\begin{array}{l}
1\ \ \ \ {\rm for\ non{\mbox{-}}derivative\ coupling,}\\
q_1\cdot q_2\ \ \ \ {\rm for\ 
 derivative\ coupling,}
\end{array}\right.
\end{equation}
and 
\begin{eqnarray}
 &&\dis{\Gamma(\phi\to f_0\gamma:m_f)=\frac{\alpha}{3}
 g_{\phi f_0\gamma}^2(m_f)\left(\frac{m_{\phi}^2-m_f^2}{2m_{\phi}}
 \right)^3},\\
 &&g_{\phi f_0\gamma}(m_f)=g_{\phi f_0\gamma}^{\rm contact}+
\frac{g_{\phi K\bar{K}}g_{f_0K\bar{K}}}{2\pi^2im_K^2}\left[\frac{2m_K^2-m_f^2}{2}
\right]I(a, b_0).
\end{eqnarray} 
In Eq. (24), factor $\dis{\left[\frac{2m_K^2-m_f^2}{2}\right]}$ is replaced to 1 
for non-derivative coupling. $g_{f_0K\bar{K}}$ is defined in the similar equation 
as Eq. (22),
$$
\hspace{2.5cm}M(f_0(q)\to K^+(q_1)+K^-(q_2))=g_{f_0K\bar{K}}\times\left\{
\begin{array}{l}
1\ \ \ \ {\rm for\ non{\mbox{-}}derivative\ coupling,}\\
q_1\cdot q_2\ \ \ \ {\rm for\ 
 derivative\ coupling,}
\end{array}\right.\hspace{2.5cm}(22')
$$
\par
We fit the Eq. (19) using the experimental data from the SND and KLEO collaboration 
in Ref. \cite{rare decay}, and obtain the best-fitted curves shown in Fig. 5; 
solid line for non-derivative coupling and dashed line for derivative one. 
\begin{figure}
\begin{center}
\includegraphics[width=12.5cm]{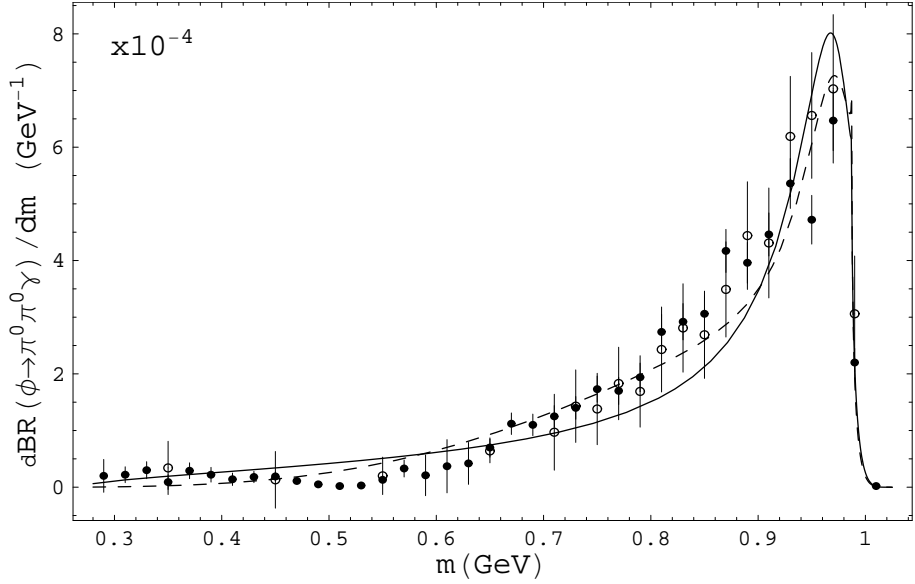}\\
\caption{$dBR(\phi\to\pi^0\pi^0\gamma)/dm$ (in unit of GeV$^{-1}$) as a function of 
the $\pi^0$-$\pi^0$ invariant mass $m$ (in GeV$^{-1}$). Solid line shows the best 
fitted curve for the non-derivative coupling and the dashed line shows the best 
fitted curve for the derivative one. 
Experimental data indicated by circles are from the SND collaboration in Ref. 
\cite{rare decay}, and those by filled circles are from KLEO collaboration 
in Ref. \cite{rare decay}.
}
\end{center}
\end{figure}
The choice of the parameters $G_1$ and $G_2$ for these best fit and estimated 
value for $BR(\phi\to\pi^0\eta\gamma)$ are obtained  as 
\begin{eqnarray}
\hspace{1cm}&G_1=7.1\times10^{-4}{\rm GeV}^{-1},\ G_2=0.001,\ &BR(\phi\to\pi^0
\pi^0\gamma)=1.06\times10^{-4}\nonumber\\ 
&&  \ {\rm for\ non\mbox{-}derivative\ coupling,}\\
&G_1=6.9\times10^{-4}{\rm GeV}^{-1},\ G_2=0.055,\ &BR(\phi\to\pi^0\pi^0\gamma)
=1.08\times10^{-4}\nonumber\\ 
&&  \ {\rm for\ derivative\ coupling.}\hspace{6.3cm}(25')\nonumber
\end{eqnarray}
Estimated value for $BR(\phi\to\pi^0\pi^0\gamma)$ is consistent with the 
experimental data $BR^{\rm exp}(\phi\to\pi^0\pi^0\gamma)=(1.09\pm0.06)\times
10^{-4}$ \cite{rare decay, PDG 2006}, and estimated value of $G_1$ is consistent 
with the value evaluated from the experimental data (Ref. \cite{PDG 2006}),
$$
G_1^{\rm exp}=\dis{\frac{2}{\pi\Gamma_f}}BR^{\rm exp}
(\phi\to f_0\gamma:m_f)BR^{\rm exp}(f_0\to \pi^0\pi^0:m_f)=(10.0\pm4.8)
\times10^{-4}{\rm GeV}^{-1}.
$$
As the case for the decay $\phi\to\pi^0\eta\gamma$, contact $g_{\phi f_0\gamma}$ 
interaction ($G_2$) is very small compared to $|I(a,b_0)|=0.783$ for $m_f=0.980$
GeV, then one can suppose that the $K^+K^-$ loop contribution is dominant in 
the $\phi\to\pi^0\pi^0\gamma$ decay. 
We suppose the $\phi\to f_0\gamma$ decay is caused from the $K^+K^-$ loop 
interaction, then can estimate the coupling constant $g_{a_0K\bar{K}}$ from the 
relation 
\begin{equation}
 \dis{\Gamma(\phi\to f_0\gamma)=\frac{\alpha}{3}
 \left|\frac{g_{\phi K\bar{K}}g_{f_0K\bar{K}}}{2\pi^2m_K^2}
 \left[\frac{2m_K^2-m_f^2}{2}\right]I(a, b_0)\right|^2\left(\frac{m_{\phi}^2-m_f^2}
 {2m_{\phi}} \right)^3},
\end{equation} 
where the factor $\dis{\left[\frac{2m_K^2-m_f^2}{2}\right]}$ is replaced to 1 for 
the non-derivative coupling. 
Using the value Eq. (13) of $g_{\phi K\bar{K}}$ and the experimental value  
$\Gamma(\phi\to f_0\gamma)=(0.323\pm0.029)\times10^{-3}{\rm MeV}$ in Ref.\cite{
PDG 2006}, we obtain the result
\begin{equation}
g_{f_0K\bar{K}}=\left\{
\begin{array}{l}
4.72\pm0.82\ \ {\rm GeV},\ \ \ {\rm for\ non\mbox{-}derivative\ coupling}\\
20.0\pm0.50\ \ {\rm GeV^{-1}}.\ \ \ {\rm for\ derivative\ coupling}
\end{array} 
\right.
\end{equation}
Using relations Eqs. (20), (21), (26) and estimated results (25), (25'), (27), 
we obtained the values for $g_{f_0\pi\pi}$,
$$
\hspace{3.8cm}g_{f_0\pi\pi}=\left\{
\begin{array}{l}
1.12\pm0.69\ \ {\rm GeV},\ \ \ \ {\rm for\ non\mbox{-}derivative\ coupling}\\
2.43\pm1.50\ \ {\rm GeV^{-1}}.\ \ \ \ {\rm for \ derivative\ coupling}
\end{array} 
\right.\hspace{3.7cm}(27')
$$
\par
The rather large value of the ratio $g_{f_0K\bar{K}}/g_{a_0K\bar{K}}\sim2$ suggests 
that the $a_0$ and $f_0$ scalar mesons are not the pure $qq\bar{q}\bar{q}$ states 
but there exist the mixing (inter-mixing) between $qq\bar{q}\bar{q}$ and $q\bar{q}$ 
scalar mesons. 
Furthermore, the existence of the coupling $g_{f_0\pi\pi}$ suggest the intra mixing 
between $qq\bar{q}\bar{q}$ $f_0(600)$ and $f_0(980)$ scalar mesons.
%
\section{Mixing between Low and High Mass Scalar Mesons}
In this section, we review the mixing among the low mass scalar, high 
mass scalar and glueball discussed in our previous work \cite{teshima1, teshima2}.
The $qq\bar{q}\bar{q}$ scalar $SU(3)$ nonet $S^b_a$ are represented by the quark 
triplet $q_a$ and anti-quark triplet $\bar{q^a}$ as
\begin{equation}
S^a_b\sim\epsilon^{acd}q_cq_d\epsilon_{bef}\bar{q}^e\bar{q}^f
\end{equation}
and have the following flavor configuration  \cite{jaffe}, \cite{mixing}:
\begin{equation}
    \begin{array}{ccc}
    \bar{d}\bar{s}su,\ \frac{1}{\sqrt{2}}(\bar{d}\bar{s}ds-\bar{s}\bar{u}su),\ 
    \bar{s}\bar{u}ds&\Longleftrightarrow&a_0^+,\ a_0^0,\ a_0^-\\
    \bar{d}\bar{s}ud,\ \bar{s}\bar{u}ud,\ \bar{u}\bar{d}su,\ \bar{u}\bar{d}ds
    &\Longleftrightarrow&\kappa^+,\ \kappa^0,\ \overline{\kappa^0},\ \kappa^-\\
    \frac{1}{\sqrt{2}}(\bar{d}\bar{s}ds+\bar{s}\bar{u}su)&\Longleftrightarrow&f_{NS}
    \sim f_0(980)\\
    \bar{u}\bar{d}ud&\Longleftrightarrow&f_{NN}\sim f_0(600) 
    \end{array} \nonumber
\end{equation}
The high mass scalar mesons $S'^a_b$ are the ordinary $SU(3)$ nonet 
$$ S'^a_b\sim \bar{q}^aq_b. $$
\par
The inter-mixing between $qq\bar{q}\bar{q}$ and $q\bar{q}$ states may be large, 
because the transition between $qq\bar{q}\bar{q}$ and $q\bar{q}$ states is 
caused by the OZI rule allowed diagram shown in fig. 6.
\begin{figure}
\vspace{0.5cm}
\begin{center}
\unitlength 0.1in
\begin{picture}(27.10,6.10)(1.60,-10.05)
%
\special{pn 13}%
\special{pa 2800 1000}%
\special{pa 1700 1000}%
\special{fp}%
\special{sh 1}%
\special{pa 1700 1000}%
\special{pa 1767 1020}%
\special{pa 1753 1000}%
\special{pa 1767 980}%
\special{pa 1700 1000}%
\special{fp}%
%
\special{pn 13}%
\special{pa 1700 1000}%
\special{pa 600 1000}%
\special{fp}%
%
\special{pn 13}%
\special{pa 600 400}%
\special{pa 1700 400}%
\special{fp}%
\special{sh 1}%
\special{pa 1700 400}%
\special{pa 1633 380}%
\special{pa 1647 400}%
\special{pa 1633 420}%
\special{pa 1700 400}%
\special{fp}%
%
\special{pn 13}%
\special{pa 1700 400}%
\special{pa 2800 400}%
\special{fp}%
%
\special{pn 8}%
\special{pa 1400 400}%
\special{pa 1400 1000}%
\special{fp}%
%
\special{pn 8}%
\special{pa 2000 1000}%
\special{pa 2000 400}%
\special{fp}%
%
\special{pn 13}%
\special{pa 1600 600}%
\special{pa 1500 600}%
\special{fp}%
\special{sh 1}%
\special{pa 1500 600}%
\special{pa 1567 620}%
\special{pa 1553 600}%
\special{pa 1567 580}%
\special{pa 1500 600}%
\special{fp}%
%
\special{pn 13}%
\special{pa 1500 600}%
\special{pa 600 600}%
\special{fp}%
%
\special{pn 13}%
\special{pa 600 800}%
\special{pa 1560 800}%
\special{fp}%
\special{sh 1}%
\special{pa 1560 800}%
\special{pa 1493 780}%
\special{pa 1507 800}%
\special{pa 1493 820}%
\special{pa 1560 800}%
\special{fp}%
%
\special{pn 13}%
\special{pa 1560 800}%
\special{pa 1600 800}%
\special{fp}%
%
\special{pn 13}%
\special{ar 1600 700 100 100  4.7123890 6.2831853}%
\special{ar 1600 700 100 100  0.0000000 1.5707963}%
\special{pn 13}%
\special{pa 1700 700}%
\special{pa 1705 693}%
\special{pa 1710 688}%
\special{pa 1715 683}%
\special{pa 1720 680}%
\special{pa 1725 679}%
\special{pa 1730 680}%
\special{pa 1735 683}%
\special{pa 1740 688}%
\special{pa 1745 693}%
\special{pa 1750 700}%
\special{pa 1755 707}%
\special{pa 1760 712}%
\special{pa 1765 717}%
\special{pa 1770 720}%
\special{pa 1775 721}%
\special{pa 1780 720}%
\special{pa 1785 717}%
\special{pa 1790 712}%
\special{pa 1795 707}%
\special{pa 1800 700}%
\special{pa 1805 693}%
\special{pa 1810 688}%
\special{pa 1815 683}%
\special{pa 1820 680}%
\special{pa 1825 679}%
\special{pa 1830 680}%
\special{pa 1835 683}%
\special{pa 1840 688}%
\special{pa 1845 693}%
\special{pa 1850 700}%
\special{pa 1855 707}%
\special{pa 1860 712}%
\special{pa 1865 717}%
\special{pa 1870 720}%
\special{pa 1875 721}%
\special{pa 1880 720}%
\special{pa 1885 717}%
\special{pa 1890 712}%
\special{pa 1895 707}%
\special{pa 1900 700}%
\special{pa 1905 693}%
\special{pa 1910 688}%
\special{pa 1915 683}%
\special{pa 1920 680}%
\special{pa 1925 679}%
\special{pa 1930 680}%
\special{pa 1935 683}%
\special{pa 1940 688}%
\special{pa 1945 693}%
\special{pa 1950 700}%
\special{pa 1955 707}%
\special{pa 1960 712}%
\special{pa 1965 717}%
\special{pa 1970 720}%
\special{pa 1975 721}%
\special{pa 1980 720}%
\special{pa 1985 717}%
\special{pa 1990 712}%
\special{pa 1995 707}%
\special{pa 2000 700}%
\special{sp}%
%
\special{pn 8}%
\special{pa 600 400}%
\special{pa 600 1000}%
\special{ip}%
%
\special{pn 4}%
\special{pa 1040 600}%
\special{pa 850 790}%
\special{fp}%
\special{pa 920 600}%
\special{pa 730 790}%
\special{fp}%
\special{pa 800 600}%
\special{pa 610 790}%
\special{fp}%
\special{pa 680 600}%
\special{pa 600 680}%
\special{fp}%
\special{pa 1160 600}%
\special{pa 970 790}%
\special{fp}%
\special{pa 1280 600}%
\special{pa 1090 790}%
\special{fp}%
\special{pa 1390 610}%
\special{pa 1210 790}%
\special{fp}%
\special{pa 1400 720}%
\special{pa 1330 790}%
\special{fp}%
%
\special{pn 4}%
\special{pa 1080 800}%
\special{pa 890 990}%
\special{fp}%
\special{pa 960 800}%
\special{pa 770 990}%
\special{fp}%
\special{pa 840 800}%
\special{pa 650 990}%
\special{fp}%
\special{pa 720 800}%
\special{pa 600 920}%
\special{fp}%
\special{pa 1200 800}%
\special{pa 1010 990}%
\special{fp}%
\special{pa 1320 800}%
\special{pa 1130 990}%
\special{fp}%
\special{pa 1400 840}%
\special{pa 1250 990}%
\special{fp}%
%
\special{pn 8}%
\special{pa 2800 400}%
\special{pa 2800 1000}%
\special{ip}%
%
\special{pn 4}%
\special{pa 2800 520}%
\special{pa 2330 990}%
\special{fp}%
\special{pa 2790 410}%
\special{pa 2210 990}%
\special{fp}%
\special{pa 2680 400}%
\special{pa 2090 990}%
\special{fp}%
\special{pa 2560 400}%
\special{pa 2000 960}%
\special{fp}%
\special{pa 2440 400}%
\special{pa 2000 840}%
\special{fp}%
\special{pa 2320 400}%
\special{pa 2000 720}%
\special{fp}%
\special{pa 2200 400}%
\special{pa 2000 600}%
\special{fp}%
\special{pa 2080 400}%
\special{pa 2000 480}%
\special{fp}%
\special{pa 2800 640}%
\special{pa 2450 990}%
\special{fp}%
\special{pa 2800 760}%
\special{pa 2570 990}%
\special{fp}%
\special{pa 2800 880}%
\special{pa 2690 990}%
\special{fp}%
\put(28.7000,-7.6000){\makebox(0,0)[lb]{$q\bar{q}$}}%
\put(1.6000,-7.6000){\makebox(0,0)[lb]{$qq\bar{q}\bar{q}$}}%
%
\special{pn 4}%
\special{pa 1120 400}%
\special{pa 930 590}%
\special{fp}%
\special{pa 1000 400}%
\special{pa 810 590}%
\special{fp}%
\special{pa 880 400}%
\special{pa 690 590}%
\special{fp}%
\special{pa 760 400}%
\special{pa 600 560}%
\special{fp}%
\special{pa 640 400}%
\special{pa 600 440}%
\special{fp}%
\special{pa 1240 400}%
\special{pa 1050 590}%
\special{fp}%
\special{pa 1360 400}%
\special{pa 1170 590}%
\special{fp}%
\special{pa 1400 480}%
\special{pa 1290 590}%
\special{fp}%
\end{picture}%
\hspace{2cm} \\
Fig. 6\ \ OZI rule allowed graph for $qq\bar{q}\bar{q}$ and $q\bar{q}$ states 
transition
\end{center}
\end{figure}
Considering the above flavor configuration for $qq\bar{q}\bar{q}$ states, the 
expression for this transition is suggested as  
\begin{eqnarray}
L_{\rm int}&=&\lambda_{01}[a_0^+{a'}_0^-+a_0^-{a'}_0^++a_0^0{a'}_0^0
    +K_0^{*+}K_0'^{*-}+K_0^{*-}K_0'^{*+}+K_0^{*0}K_0'^{*0}+\bar{K_0^*}^0
    \bar{K'}_0^{*0}      \nonumber\\
    &&+\sqrt{2}f_{NN}f'_N+f_{NS}f'_N+\sqrt{2}f_{NS}f'_S].
\end{eqnarray}
The parameter $\lambda_{01}$ represents the strength of the inter-mixing and can 
be considered as rather large. 
When we represent the $I=1$ pure $qq\bar{q}\bar{q}$ and $q\bar{q}$ states by 
$\overline{a_0(980)}$ and $\overline{a_0(1450)}$, and masses for these states by
$m^2_{\overline{a_0(980)}}$ and $m^2_{\overline{a_0(1450)}}$, the mass matrix is 
represented as 
\begin{equation}
\left(\begin{array}{cc}
    m^2_{\overline{a_0(980)}} & \lambda_{01}\\
    \lambda_{01} & m^2_{\overline{a_0(1450)}}
    \end{array}\right).
\end{equation}
Diagonalising this mass matrix, we can get the masses for the physical states 
$a_0(980)$ and $a_0(1450)$ represented as mixing states of $\overline{a_0(980)}$ 
and $\overline{a_0(1450)}$; 
\begin{equation}
\begin{array}{l}
a_0(980)=\cos\theta_a\overline{a_0(980)}-\sin\theta_a\overline{a_0(1450)},\\
a_0(1450)=\sin\theta_a\overline{a_0(980)}+\cos\theta_a\overline{a_0(1450)}.
\end{array}
\end{equation} 
Mixing angle $\theta_a$ and before mixing state masses $m_{\overline
{a_0(980)}}$ and $m_{\overline{a_0(1450)}}$ are represented by the 
inter-mixing parameter $\lambda_{01}$ as 
\begin{eqnarray}
&&\epsilon_a=\frac{m_{a_0(1450)}^2-m_{a_0(980)}^2}{2}-\sqrt{\left(
\frac{m_{a_0(1450)}^2-m_{a_0(980)}^2}{2}\right)^2-\lambda_{01}^2},\nonumber\\
&&\theta_a=\tan^{-1}\frac{\epsilon_a}{\lambda_{01}}\\
&&m_{\overline{a_0(1450)}}=\sqrt{m_{a_0(1450)}^2+\epsilon_a},\ \ \ 
m_{\overline{a_0(980)}}=\sqrt{m_{a_0(980)}^2-\epsilon_a},\nonumber
\end{eqnarray}
where $m_{a_0(980)}$ and $m_{a_0(1450)}$ are the masses of the states 
$a_0(980)$ and $a_0(1450)$.
\par
Similarly, for the $I=1/2$ $K_0^*(800)$ and $K_0^*(1430)$ mesons, the mass matrix  
is represented as 
\begin{equation}
\left(\begin{array}{cc}
    m^2_{\overline{K_0^*(800)}} & \lambda_{01}\\
    \lambda_{01} & m^2_{\overline{K_0^*(1430)}}
    \end{array}\right), 
\end{equation}
where $m_{\overline{K_0^*(800)}}$ and $m_{\overline{K_0^*(1430)}}$ are the masses 
of pure $qq\bar{q}\bar{q}$ and $q\bar{q}$ states  $\overline{K_0^*(800)}$ and 
$\overline{K_0^*(1430)}$.
The physical states $K_0^*(800)$ and $K_0^*(1430)$ are written by the 
before mixing states $\overline{K_0^*(800)}$ and $\overline{K_0^*(1430)}$ and 
mixing angle $\theta_{K}$ as 
\begin{equation}
\begin{array}{l}
K_0^*(800)=\cos\theta_K\overline{K_0^*(800)}-\sin\theta_K\overline{K_0^*(1430)},\\
K_0^*(1430)=\sin\theta_K\overline{K_0^*(800)}+\cos\theta_K\overline{K_0^*(1430)}.
\end{array}
\end{equation}  
Mixing angle $\theta_K$ and before mixing state masses $m_{\overline
{K_0^*(800)}}$ and $m_{\overline{K_0^*(1430)}}$ are represented by the 
inter-mixing parameter $\lambda_{01}$ as 
\begin{eqnarray}
&&\epsilon_K=\frac{m_{K_0^*(1430)}^2-m_{K_0^*(800)}^2}{2}-\sqrt{\left(
\frac{m_{K_0^*(1430)}^2-m_{K_0^*(800)}^2}{2}\right)^2-\lambda_{01}^2},\nonumber\\
&&\theta_K=\tan^{-1}\frac{\epsilon_K}{\lambda_{01}}\\
&&m_{\overline{K_0^*(1430)}}=\sqrt{m_{K_0^*(1430)}^2+\epsilon_K},\ \ \ 
m_{\overline{K_0^*(800)}}=\sqrt{m_{K_0^*(800)}^2-\epsilon_K},\nonumber
\end{eqnarray}
where $m_{K_0^*(800)}$ and $m_{K_0^*(1430)}$ are the masses of the physical 
states $K_0^*(800)$ and $K_0^*(1430)$. 
\par
Next, we consider the mixing between $I=0$ low and high mass scalar mesons.  
Among the $I=0, L=1\ q\bar{q}$ scalar mesons, there are the intra-mixing weaker 
than the inter-mixing, caused from the transition between themselves represented by 
the OZI rule suppression graph shown in Fig. 7, and furthermore the mixing 
between the $q\bar{q}$ scalar meson and the glueball caused from the transition 
represented by the graph shown in Fig. 8(a). 
\begin{figure}
\vspace{0.5cm}
\begin{center}
\input {fig07.tex}\\
Fig. 7  OZI rule suppression graph for $q\bar{q}-q\bar{q}$ transition.
\end{center}
\end{figure}
\begin{figure}
\begin{center}
\input{fig08.tex}
\vspace{0.5cm}\\ 
Fig. 8 Transition graph between (a) $q\bar{q}$ and $gg$, and (b) $gg$ and $gg$.
\end{center}
\end{figure}
Thus, the mass matrix for these $I=0, L=1\ q\bar{q}$ 
scalar mesons and glueball is represented as 
\begin{eqnarray}
&&\left(\begin{array}{ccc}
m^2_{N'}+2\lambda_1&\sqrt{2}\lambda_1&\sqrt{2}\lambda_G\\
\sqrt{2}\lambda_1&m^2_{S'}+\lambda_1&\lambda_G\\ 
\sqrt{2}\lambda_G&\lambda_G&\lambda_{GG}
\end{array}\right),\\
&& m^2_{N'}=m^2_{\overline{a_0(1450)}}, \ \ m^2_{S'}=2m^2_{\overline{K^*_0(1430)}}-
m^2_{\overline{a_0(1450)}}, \nonumber
\end{eqnarray}
where $\lambda_1$ is the transition strength among the $I=0\ q\bar{q}$ mesons, 
$\lambda_{G}$ is the transition strength between $q\bar{q}$ and glueball $gg$ and 
$\lambda_{GG}$ is the pure glueball mass square. For the light $I=0\ qq\bar{q}\bar{q}$ scalar mesons, there are the intra-mixing 
caused from the transition between themselves represented by the OZI rule 
suppression graph shown in Fig. 9, and the mass matrix for these $I=0\ qq\bar{q}
\bar{q}$ scalar meson is represented as  
\begin{eqnarray}
&&\left(\begin{array}{cc}
m^2_{NN}+\lambda_0&\sqrt{2}\lambda_0\\
\sqrt{2}\lambda_0&m^2_{NS}+2\lambda_0 
\end{array}\right), \\
&&m^2_{NN}=2m^2_{\overline{K^*_0(800)}}-m^2_{\overline{a_0(980)}},\ \ \ m^2_{NS}=
m^2_{\overline{a_0(980)}}, \nonumber
\end{eqnarray}
where $\lambda_0$ represents the transition strength between $I=0\ \ qq\bar{q}
\bar{q}$ mesons. 
\begin{figure}
\begin{center}
\unitlength 0.1in
\begin{picture}(25.00,6.10)(2.00,-10.05)
%
\special{pn 13}%
\special{pa 600 400}%
\special{pa 1600 400}%
\special{fp}%
\special{sh 1}%
\special{pa 1600 400}%
\special{pa 1533 380}%
\special{pa 1547 400}%
\special{pa 1533 420}%
\special{pa 1600 400}%
\special{fp}%
%
\special{pn 13}%
\special{pa 1600 400}%
\special{pa 2600 400}%
\special{fp}%
%
\special{pn 13}%
\special{pa 2600 1000}%
\special{pa 1600 1000}%
\special{fp}%
\special{sh 1}%
\special{pa 1600 1000}%
\special{pa 1667 1020}%
\special{pa 1653 1000}%
\special{pa 1667 980}%
\special{pa 1600 1000}%
\special{fp}%
%
\special{pn 13}%
\special{pa 1600 1000}%
\special{pa 600 1000}%
\special{fp}%
%
\special{pn 8}%
\special{pa 1200 1000}%
\special{pa 1200 400}%
\special{fp}%
%
\special{pn 8}%
\special{pa 2000 400}%
\special{pa 2000 1000}%
\special{fp}%
%
\special{pn 8}%
\special{pa 2600 1000}%
\special{pa 2600 400}%
\special{ip}%
%
\special{pn 8}%
\special{pa 600 400}%
\special{pa 600 1000}%
\special{ip}%
%
\special{pn 4}%
\special{pa 1120 400}%
\special{pa 600 920}%
\special{fp}%
\special{pa 1200 440}%
\special{pa 650 990}%
\special{fp}%
\special{pa 1200 560}%
\special{pa 770 990}%
\special{fp}%
\special{pa 1200 680}%
\special{pa 890 990}%
\special{fp}%
\special{pa 1200 800}%
\special{pa 1010 990}%
\special{fp}%
\special{pa 1200 920}%
\special{pa 1130 990}%
\special{fp}%
\special{pa 1000 400}%
\special{pa 600 800}%
\special{fp}%
\special{pa 880 400}%
\special{pa 600 680}%
\special{fp}%
\special{pa 760 400}%
\special{pa 600 560}%
\special{fp}%
\special{pa 640 400}%
\special{pa 600 440}%
\special{fp}%
%
\special{pn 4}%
\special{pa 2560 400}%
\special{pa 2000 960}%
\special{fp}%
\special{pa 2600 480}%
\special{pa 2090 990}%
\special{fp}%
\special{pa 2600 600}%
\special{pa 2210 990}%
\special{fp}%
\special{pa 2600 720}%
\special{pa 2330 990}%
\special{fp}%
\special{pa 2600 840}%
\special{pa 2450 990}%
\special{fp}%
\special{pa 2440 400}%
\special{pa 2000 840}%
\special{fp}%
\special{pa 2320 400}%
\special{pa 2000 720}%
\special{fp}%
\special{pa 2200 400}%
\special{pa 2000 600}%
\special{fp}%
\special{pa 2080 400}%
\special{pa 2000 480}%
\special{fp}%
%
\special{pn 13}%
\special{pa 600 600}%
\special{pa 1300 600}%
\special{fp}%
\special{sh 1}%
\special{pa 1300 600}%
\special{pa 1233 580}%
\special{pa 1247 600}%
\special{pa 1233 620}%
\special{pa 1300 600}%
\special{fp}%
%
\special{pn 13}%
\special{pa 1300 800}%
\special{pa 1200 800}%
\special{fp}%
\special{sh 1}%
\special{pa 1200 800}%
\special{pa 1267 820}%
\special{pa 1253 800}%
\special{pa 1267 780}%
\special{pa 1200 800}%
\special{fp}%
%
\special{pn 13}%
\special{pa 1200 800}%
\special{pa 600 800}%
\special{fp}%
%
\special{pn 13}%
\special{pa 1900 600}%
\special{pa 2000 600}%
\special{fp}%
\special{sh 1}%
\special{pa 2000 600}%
\special{pa 1933 580}%
\special{pa 1947 600}%
\special{pa 1933 620}%
\special{pa 2000 600}%
\special{fp}%
%
\special{pn 13}%
\special{pa 2000 600}%
\special{pa 2600 600}%
\special{fp}%
%
\special{pn 13}%
\special{pa 2600 800}%
\special{pa 1900 800}%
\special{fp}%
\special{sh 1}%
\special{pa 1900 800}%
\special{pa 1967 820}%
\special{pa 1953 800}%
\special{pa 1967 780}%
\special{pa 1900 800}%
\special{fp}%
%
\special{pn 13}%
\special{ar 1300 700 100 100  4.7123890 6.2831853}%
\special{ar 1300 700 100 100  0.0000000 1.5707963}%
%
\special{pn 13}%
\special{ar 1900 700 100 100  1.5707963 4.7123890}%
\special{pn 13}%
\special{pa 1400 700}%
\special{pa 1405 693}%
\special{pa 1410 686}%
\special{pa 1415 679}%
\special{pa 1420 674}%
\special{pa 1425 671}%
\special{pa 1430 669}%
\special{pa 1435 668}%
\special{pa 1440 670}%
\special{pa 1445 673}%
\special{pa 1450 677}%
\special{pa 1455 683}%
\special{pa 1460 690}%
\special{pa 1465 698}%
\special{pa 1470 705}%
\special{pa 1475 712}%
\special{pa 1480 719}%
\special{pa 1485 724}%
\special{pa 1490 728}%
\special{pa 1495 731}%
\special{pa 1500 732}%
\special{pa 1505 731}%
\special{pa 1510 728}%
\special{pa 1515 724}%
\special{pa 1520 719}%
\special{pa 1525 712}%
\special{pa 1530 705}%
\special{pa 1535 698}%
\special{pa 1540 690}%
\special{pa 1545 683}%
\special{pa 1550 677}%
\special{pa 1555 673}%
\special{pa 1560 670}%
\special{pa 1565 668}%
\special{pa 1570 669}%
\special{pa 1575 671}%
\special{pa 1580 674}%
\special{pa 1585 679}%
\special{pa 1590 686}%
\special{pa 1595 693}%
\special{pa 1600 700}%
\special{pa 1605 707}%
\special{pa 1610 714}%
\special{pa 1615 721}%
\special{pa 1620 726}%
\special{pa 1625 729}%
\special{pa 1630 731}%
\special{pa 1635 732}%
\special{pa 1640 730}%
\special{pa 1645 727}%
\special{pa 1650 723}%
\special{pa 1655 717}%
\special{pa 1660 710}%
\special{pa 1665 702}%
\special{pa 1670 695}%
\special{pa 1675 688}%
\special{pa 1680 681}%
\special{pa 1685 676}%
\special{pa 1690 672}%
\special{pa 1695 669}%
\special{pa 1700 668}%
\special{pa 1705 669}%
\special{pa 1710 672}%
\special{pa 1715 676}%
\special{pa 1720 681}%
\special{pa 1725 688}%
\special{pa 1730 695}%
\special{pa 1735 702}%
\special{pa 1740 710}%
\special{pa 1745 717}%
\special{pa 1750 723}%
\special{pa 1755 727}%
\special{pa 1760 730}%
\special{pa 1765 732}%
\special{pa 1770 731}%
\special{pa 1775 729}%
\special{pa 1780 726}%
\special{pa 1785 721}%
\special{pa 1790 714}%
\special{pa 1795 707}%
\special{pa 1800 700}%
\special{sp}%
\put(2.0000,-7.5000){\makebox(0,0)[lb]{$qq\bar{q}\bar{q}$}}%
\put(27.0000,-7.5000){\makebox(0,0)[lb]{$qq\bar{q}\bar{q}$}}%
\end{picture}%
\vspace{0.5cm}\\
Fig. 9. OZI suppression graph for $qq\bar{q}\bar{q}-qq\bar{q}\bar{q}$ transition.
\end{center}
\end{figure}
\par
The inter- and intra-mixing among $I=0$ low mass and high mass scalar mesons and 
glueball is expressed by the overall mixing mass matrix as
\begin{equation}
\left(\begin{array}{ccccc}
    m^2_{NN}+\lambda_0&\sqrt{2}\lambda_0&\sqrt{2}\lambda_{01}&0&0\\
    \sqrt{2}\lambda_0&m^2_{NS}+2\lambda_0&\lambda_{01}&\sqrt{2}\lambda_{01}&0\\
    \sqrt{2}\lambda_{01}&\lambda_{01}&m^2_{N'}+2\lambda_1&\sqrt{2}\lambda_1&
    \sqrt{2}\lambda_{G}\\
    0&\sqrt{2}\lambda_{01}&\sqrt{2}\lambda_1&m^2_{S'}+\lambda_1&
    \lambda_{G}\\
    0&0&\sqrt{2}\lambda_{G}&\lambda_{G}&\lambda_{GG}
    \end{array}\right).
\end{equation}
Diagonalising this mass matrix, we obtain the eigenvalues of low mass 
and high mass scalar mesons $I=0$ states $f_0(600)$, $f_0(980)$, $f_0(1370)$, 
$f_0(1500)$ and $f_0(1710)$. The eigenstates of these scalar mesons are 
represented as follows;
\begin{eqnarray}
&&\left(\begin{array}{c}
    f_0(600)\\
    f_0(980)\\
    f_0(1370)\\
    f_0(1500)\\
    f_0(1710)
    \end{array}\right)=[R_{f_0(M)I}]  
    \left(\begin{array}{c}
    f_{NN}\\
    f_{NS}\\
    f_{N'}\\
    f_{S'}\\
    f_G
    \end{array}\right),\hspace{5cm}\\    
&&  [R_{f_0(M)I}]=
  \left(\begin{array}{ccccc}
     R_{f_0(600)NN}&R_{f_0(600)NS}&R_{f_0(600)N'}&R_{f_0(600)S'}&R_{f_0(600)G}\\
     R_{f_0(980)NN}&R_{f_0(980)NS}&R_{f_0(980)N'}&R_{f_0(980)S'}&R_{f_0(980)G}\\
     R_{f_0(1370)NN}&R_{f_0(1370)NS}&R_{f_0(1370)N'}&R_{f_0(1370)S'}&
     R_{f_0(1370)G}\\
     R_{f_0(1500)NN}&R_{f_0(1500)NS}&R_{f_0(1500)N'}&R_{f_0(1500)S'}&
     R_{f_0(1500)G}\\
     R_{f_0(1710)NN}&R_{f_0(1710)NS}&R_{f_0(1710)N'}&R_{f_0(1710)S'}&
     R_{f_0(1710)G}     
     \end{array}\right).\nonumber
\end{eqnarray}
\par
Using the inter-mixing parameter $\lambda_{01}$ and the mass values, $m_{a_0(9
80)}=(0.9848\pm0.0012){\rm GeV}$, $m_{a_0(1450)}=(1.474\pm0.019){\rm GeV}$, 
$m_{K_0^*(800)}=(0.841\pm0.030){\rm GeV}$ and $m_{K_0^*(1430)}=(1.414\pm0.006)
{\rm GeV}$, we obtained the mixing angles $\theta_a$, $\theta_K$ and before mixing 
states masses $m_{\overline{a_0(980)}}$, $m_{\overline{a_0(1450)}}$, $m_{
\overline{K_0^*(800)}}$, $m_{\overline{K_0^*(1430)}}$ from the relations (32) and 
(35). 
Using the values of $m^2_{N'}$, $m^2_{S'}$, $m^2_{NN}$ and $m^2_{NS}$ obtained 
from the second equations in Eqs.\,(36) and (37), and parameters 
$\lambda_0$, $\lambda_1$, $\lambda_G$ and $\lambda_{GG}$, we diagonalize the mass 
matrix Eq\,(38). 
When we fit the eigenvalues obtained to the following experimental mass 
values \cite{PDG 2006},
\begin{equation}
\begin{array}{l}
    m_{f_0(600)}=0.80\pm0.40{\rm GeV},\ m_{f_0(980)}=0.980\pm0.010{\rm GeV}, \\  
    m_{f_0(1370)}=1.350\pm0.150{\rm GeV},\ m_{f_0(1500)}=1.507\pm0.005{\rm GeV}, \\
    m_{f_0(1710)}=1.718\pm0.006{\rm GeV}, 
\end{array}    
\end{equation}
we obtain the allowed values for $\lambda_0$, $\lambda_1$, $\lambda_G$ and 
$\lambda_{GG}$. 
We tabulated the $\theta_a$, $\theta_K$, $\lambda_0$, $\lambda_1$, 
$\lambda_G$, $\lambda_{GG}$, and $R_{f_0(980)NN}$, $R_{f_0(980)NS}$, $R_{f_0(980)N'}$, 
$R_{f_0(980)S'}$, $R_{f_0(980)G}$ for the various values of $\lambda_{01}$ in the 
Table I. 
\begin{table}
\begin{footnotesize}
\begin{tabular}{|c||c|c|c|c|c|c|}
\hline
 $\lambda_{01}({\rm GeV}^2)$     &$\theta_a(^\circ)$     &$\theta_K(^\circ)$     
&$\lambda_0({\rm GeV}^2)$     &$\lambda_1({\rm GeV}^2)$     &$\lambda_G({\rm GeV}^2)$
&$\lambda_{GG}({\rm GeV}^2)$     \\
\cline{2-7}
     &$R_{f_0(980)NN}$&$R_{f_0(980)NS}$&$R_{f_0(980)N'}$&$R_{f_0(980)S'}$
     &$R_{f_0(980)G}$&     \\
\hline
$0.20$&$9.7\pm0.5$     &$9.0\pm0.5$     &$0.018\pm0.009$     &$0.275\pm0.007$     
&$0.04\pm0.04$     &$(1.152\pm0.008)^2$     \\
\cline{2-7}
     &$-0.023\pm0.014$&$-0.972\pm0.002$&$0.065\pm0.006$&$0.226\pm0.004$
     &$-0.010\pm0.010$&     \\
     \hline
$0.25$&$12.3\pm0.6$     &$11.4\pm0.6$     &$0.032\pm0.010$     &$0.264\pm0.008$     
&$0.05\pm0.05$     &$(1.512\pm0.007)^2$     \\
\cline{2-7}
     &$-0.027\pm0.026$&$-0.954\pm0.003$&$0.086\pm0.008$&$0.284\pm0.005$
     &$-0.016\pm0.016$&     \\
     \hline
$0.30$&$15.0\pm0.8$     &$13.8\pm0.8$     &$0.050\pm0.009$     &$0.252\pm0.009$     
&$0.04\pm0.04$     &$(1.512\pm0.008)^2$     \\
\cline{2-7}
     &$-0.046\pm0.024$&$-0.932\pm0.004$&$0.110\pm0.009$&$0.341\pm0.006$
     &$-0.016\pm0.016$&     \\
     \hline
$0.35$&$17.8\pm1.0$     &$16.4\pm1.0$     &$0.072\pm0.012$     &$0.233\pm0.008$
     &$0.05\pm0.05$     &$(1.511\pm0.008)^2$     \\
\cline{2-7}
     &$-0.065\pm0025$&$-0.902\pm0.007$&$0.140\pm0.012$&$0.401\pm0.007$
     &$-0.024\pm0.024$&     \\
     \hline
$0.40$&$20.8\pm1.2$     &$19.1\pm1.2$     &$0.104\pm0.012$     &$0.213\pm0.009$
     &$0.05\pm0.05$     &$(1.509\pm0.006)^2$     \\
\cline{2-7}
     &$-0.094\pm0.021$&$-0.864\pm0.010$&$0.178\pm0.014$&$0.461\pm0.007$
     &$-0.028\pm0.028$&     \\
     \hline
$0.45$&$24.2\pm1.6$     &$22.1\pm1.5$     &$0.146\pm0.014$     &$0.178\pm0.007$
     &$0.04\pm0.04$     &$(1.506\pm0.002)^2$     \\
\cline{2-7}
     &$-0.116\pm0.021$&$-0.813\pm0.011$&$0.226\pm0.015$&$0.523\pm0.006$
     &$-0.014\pm0.014$&     \\
     \hline
\end{tabular}
\end{footnotesize}
\caption{The values of mixing angles $\theta_a$, $\theta_K$, and the transition 
parameters $\lambda_0$, $\lambda_1$, $\lambda_G$  $\lambda_{GG}$, and mixing 
parameters $R_{f_0(980)NN}$, $R_{f_0(980)NS}$, $R_{f_0(980)N'}$, $R_{f_0(980)S'}$, 
$R_{f_0(980)G}$ for the various values of $\lambda_{01}$.}
\end{table}
%
\section{Coupling constant $g_{SPP}$ and mixing between $qq\bar{q}\bar{q}$ and  
$q\bar{q}$ scalar mesons}
In this section, we first express the $g_{SPP}$'s by the mixing angle $\theta_a$, 
$\theta_K$ and mixing parameters $R_{f_0NS}$ etc. 
Next, we obtain the values of the $g_{SPP}$ using the various $S\to PP$ decay 
widths and compare these values with the ones obtained from $\phi$ decay, 
and then suggest the importance of the mixing between $qq\bar{q}\bar{q}$ and 
$q\bar{q}$ scalar mesons. 
\par
We use the following expressions for $S(qq\bar{q}\bar{q}\ {\rm scalar\ meson})PP$, 
$S'(q\bar{q}\ {\rm scalar\ meson})PP$ and $G({\rm pure\ glueball})PP$ coupling with 
coupling constants $A$, $A'$ and $A''$, respectively \cite{mixing, teshima2}, 
\begin{eqnarray}
\hspace{1.5cm}L_I&=&A\varepsilon^{abc}\varepsilon_{def}S^d_a{P}^e_b{P}^f_c+
A'S'^b_a\{{P}^c_b,\ {P}^a_c\}+A''G\{{P}^b_a,\ {P}^a_b\}, \nonumber\\
&&\hspace{5cm}{\rm for\ non\mbox{-}derivative\ coupling}\\
&&A\varepsilon^{abc}\varepsilon_{def}S^d_a\partial^\mu{P}^e_b\partial_\mu{P}^f_c+
A'S'^b_a\{\partial^\mu{P}^c_b,\ \partial_\mu{P}^a_c\}+A''G\{\partial^\mu{P}^b_a,
\ \partial_\mu{P}^a_b\}.\nonumber\\
&&\hspace{5cm}{\rm for\ derivative\ coupling}\hspace{4cm}(41')\nonumber
\end{eqnarray}
These interactions are represented graphically by the diagrams shown in Fig.~10. 
\begin{figure}
\begin{center}
\unitlength 0.1in
\begin{picture}( 17.5000, 11.0700)(  2.6000,-16.7900)
%
\special{pn 13}%
\special{pa 394 888}%
\special{pa 1184 888}%
\special{fp}%
\special{pa 1184 888}%
\special{pa 1976 572}%
\special{fp}%
\special{pa 394 1364}%
\special{pa 1184 1364}%
\special{fp}%
\special{pa 1184 1364}%
\special{pa 1976 1680}%
\special{fp}%
\special{pa 1976 1680}%
\special{pa 1976 1680}%
\special{fp}%
%
\special{pn 13}%
\special{pa 394 1206}%
\special{pa 1026 1206}%
\special{fp}%
\special{pa 1026 1206}%
\special{pa 1976 826}%
\special{fp}%
\special{pa 394 1048}%
\special{pa 1026 1048}%
\special{fp}%
\special{pa 1026 1048}%
\special{pa 1184 1110}%
\special{fp}%
\special{pa 1248 1142}%
\special{pa 1976 1442}%
\special{fp}%
%
\special{pn 13}%
\special{pa 394 888}%
\special{pa 710 888}%
\special{fp}%
\special{sh 1}%
\special{pa 710 888}%
\special{pa 644 868}%
\special{pa 658 888}%
\special{pa 644 908}%
\special{pa 710 888}%
\special{fp}%
%
\special{pn 13}%
\special{pa 386 1048}%
\special{pa 702 1048}%
\special{fp}%
\special{sh 1}%
\special{pa 702 1048}%
\special{pa 636 1028}%
\special{pa 650 1048}%
\special{pa 636 1068}%
\special{pa 702 1048}%
\special{fp}%
%
\special{pn 13}%
\special{pa 1026 1206}%
\special{pa 648 1206}%
\special{fp}%
\special{sh 1}%
\special{pa 648 1206}%
\special{pa 714 1226}%
\special{pa 700 1206}%
\special{pa 714 1186}%
\special{pa 648 1206}%
\special{fp}%
%
\special{pn 13}%
\special{pa 964 1364}%
\special{pa 648 1364}%
\special{fp}%
\special{sh 1}%
\special{pa 648 1364}%
\special{pa 714 1384}%
\special{pa 700 1364}%
\special{pa 714 1344}%
\special{pa 648 1364}%
\special{fp}%
%
\special{pn 13}%
\special{pa 1184 888}%
\special{pa 1580 730}%
\special{fp}%
\special{sh 1}%
\special{pa 1580 730}%
\special{pa 1512 736}%
\special{pa 1530 750}%
\special{pa 1526 774}%
\special{pa 1580 730}%
\special{fp}%
%
\special{pn 13}%
\special{pa 1248 1142}%
\special{pa 1572 1276}%
\special{fp}%
\special{sh 1}%
\special{pa 1572 1276}%
\special{pa 1518 1232}%
\special{pa 1524 1256}%
\special{pa 1504 1270}%
\special{pa 1572 1276}%
\special{fp}%
%
\special{pn 13}%
\special{pa 1896 858}%
\special{pa 1540 1000}%
\special{fp}%
\special{sh 1}%
\special{pa 1540 1000}%
\special{pa 1610 994}%
\special{pa 1590 980}%
\special{pa 1596 956}%
\special{pa 1540 1000}%
\special{fp}%
%
\special{pn 13}%
\special{pa 1874 1640}%
\special{pa 1534 1498}%
\special{fp}%
\special{sh 1}%
\special{pa 1534 1498}%
\special{pa 1588 1542}%
\special{pa 1582 1518}%
\special{pa 1602 1504}%
\special{pa 1534 1498}%
\special{fp}%
\put(20.1000,-7.9000){\makebox(0,0)[lb]{$P$}}%
\put(20.1000,-16.4000){\makebox(0,0)[lb]{$P$}}%
\put(2.6000,-12.0000){\makebox(0,0)[lb]{$S$}}%
\end{picture}%
\hspace{1cm}
\unitlength 0.1in
\begin{picture}( 18.8000,  9.4100)(  2.2000,-15.6100)
%
\special{pn 13}%
\special{pa 364 962}%
\special{pa 1220 962}%
\special{fp}%
\special{pa 1220 962}%
\special{pa 2076 620}%
\special{fp}%
\special{pa 364 1220}%
\special{pa 1220 1220}%
\special{fp}%
\special{pa 1220 1220}%
\special{pa 2076 1562}%
\special{fp}%
%
\special{pn 13}%
\special{pa 2076 1348}%
\special{pa 1392 1082}%
\special{fp}%
\special{pa 1392 1082}%
\special{pa 2076 826}%
\special{fp}%
%
\special{pn 13}%
\special{pa 364 962}%
\special{pa 792 962}%
\special{fp}%
\special{sh 1}%
\special{pa 792 962}%
\special{pa 726 942}%
\special{pa 740 962}%
\special{pa 726 982}%
\special{pa 792 962}%
\special{fp}%
%
\special{pn 13}%
\special{pa 964 1220}%
\special{pa 724 1220}%
\special{fp}%
\special{sh 1}%
\special{pa 724 1220}%
\special{pa 790 1240}%
\special{pa 776 1220}%
\special{pa 790 1200}%
\special{pa 724 1220}%
\special{fp}%
%
\special{pn 13}%
\special{pa 1220 962}%
\special{pa 1734 758}%
\special{fp}%
\special{sh 1}%
\special{pa 1734 758}%
\special{pa 1664 764}%
\special{pa 1684 778}%
\special{pa 1680 800}%
\special{pa 1734 758}%
\special{fp}%
%
\special{pn 13}%
\special{pa 2076 834}%
\special{pa 1768 938}%
\special{fp}%
\special{sh 1}%
\special{pa 1768 938}%
\special{pa 1838 936}%
\special{pa 1818 920}%
\special{pa 1824 898}%
\special{pa 1768 938}%
\special{fp}%
%
\special{pn 13}%
\special{pa 2076 1562}%
\special{pa 1690 1408}%
\special{fp}%
\special{sh 1}%
\special{pa 1690 1408}%
\special{pa 1744 1450}%
\special{pa 1740 1428}%
\special{pa 1760 1414}%
\special{pa 1690 1408}%
\special{fp}%
%
\special{pn 13}%
\special{pa 1520 1126}%
\special{pa 1810 1246}%
\special{fp}%
\special{sh 1}%
\special{pa 1810 1246}%
\special{pa 1756 1202}%
\special{pa 1762 1226}%
\special{pa 1742 1238}%
\special{pa 1810 1246}%
\special{fp}%
\put(21.0000,-8.1000){\makebox(0,0)[lb]{$P$}}%
\put(20.9000,-15.2000){\makebox(0,0)[lb]{$P$}}%
\put(2.2000,-11.9000){\makebox(0,0)[lb]{$S'$}}%
\end{picture}%
\hspace{1cm}\input{fig103.tex}\\
Fig 10. $SPP$, $S'PP$ and $GPP$ coupling
\end{center}
\end{figure}
We define the coupling constants $g_{SPP'}$ in the 
following expression, 
\begin{eqnarray}
L_I&=&g_{a_0K\overline{K}}[\partial^\mu]\overline{K}\bi{\tau}\bi{\cdot}\bi{a_0}
[\partial_\mu]{K}
+g_{a'_0K\overline{K}}[\partial^\mu]\overline{K}\bi{\tau}\bi{\cdot}\bi{a'_0}
[\partial_\mu]{K}+
g_{a_0\pi\eta}\bi{a_0\cdot}[\partial^\mu]\bi{\pi}[\partial_\mu]\eta\nonumber\\
&+&g_{a'_0\pi\eta}\bi{a'_0\cdot}[\partial^\mu]\bi{\pi}[\partial_\mu]\eta
+g_{a_0\pi\eta'}\bi{a_0\cdot}[\partial^\mu]\bi{\pi}[\partial_\mu]\eta'
+g_{a'_0\pi\eta'}\bi{a'_0\cdot}[\partial^\mu]\bi{\pi}[\partial_\mu]\eta'\nonumber\\
&+&g_{K^*_0 K\pi}([\partial^\mu]\overline{K}\bi{\tau\cdot}[\partial_\mu]\bi{\pi}
K_0^*+H.C.)
+g_{{K^*_0}'K\pi}([\partial^\mu]\overline{K}\bi{\tau\cdot}[\partial_\mu]\bi{\pi}
{K^*_0}'+H.C.)\nonumber\\
&+&g_{K^*_0K\eta}(\overline{K^*_0}[\partial^\mu]K[\partial_\mu]\eta+H.C.)
+g_{{K^*_0}'K\eta}(\overline{{K^*_0}'}[\partial^\mu]K[\partial_\mu]\eta+H.C.)
\nonumber\\
&+&g_{K^*_0 K\eta'}(\overline{K^*_0}[\partial^\mu]K[\partial_\mu]\eta'+H.C.)
+g_{{K^*_0}'K\eta'}(\overline{{K^*_0}'}[\partial^\mu]K[\partial_\mu]\eta'+H.C.),
\nonumber\\
&+&g_{f_0(M)\pi\pi}\frac12f_0(M)[\partial^\mu]\bi{\pi}{\cdot}[\partial_\mu]\bi{\pi}
+g_{f_0(M)K\overline{K}}f_0(M)[\partial^\mu]K[\partial_\mu]\overline{K}
+g_{f_0(M)\eta\eta}f_0(M)[\partial^\mu]\eta[\partial_\mu]\eta
\nonumber\\
&+&g_{f_0(M)\eta\eta'}f_0(M)[\partial^\mu]\eta[\partial_\mu]\eta'
+g_{f_0(M)\eta'\eta'}f_0(M)[\partial^\mu]\eta'[\partial_\mu]\eta',
\end{eqnarray}
where $[\partial^\mu]$'s are replaced to 1 for non-derivative couplings. 
These definitions of $g_{SPP}$'s are the same as ones of $\gamma_{SPP}$ in our 
previous work \cite{teshima2} except for $g_{a_0K\bar{K}}$, $g_{K^*_0 K\pi}$, 
which are related as $\sqrt{2}g_{a_0K\bar{K}}=\gamma_{a_0K\bar{K}}$, 
$\sqrt{2}g_{K^*_0 K\pi}=\gamma_{K^*_0 K\pi}$. 
\par 
Then the  coupling constants $g_{SPP}$'s are expressed  as
\begin{eqnarray}
&&g_{a_0(980)K\bar{K}}=\sqrt{2}(A\cos\theta_a-A'\sin{\theta_a}),\nonumber\\
&&g_{a_0(1450)K\bar{K}}=\sqrt{2}(A\sin\theta_a+A'\cos{\theta_a}),\nonumber\\
&&g_{a_0(980)\pi\eta}=2(A\cos{\theta_a}\sin{\theta_P}-\sqrt{2}A'
\sin{\theta_a}\cos{\theta_P}) ,\nonumber\\
&&g_{a_0(1450)\pi\eta}=2(A\sin{\theta_a}\sin{\theta_P}+\sqrt{2}A'
\cos{\theta_a}\cos{\theta_P}) ,\nonumber\\
&&g_{a_0(1450)\pi\eta'}=2(-A\sin{\theta_a}\cos{\theta_P}+\sqrt{2}A'
\cos{\theta_a}\sin{\theta_P}) ,\nonumber\\
&&g_{K^*_0(800)\pi K}=\sqrt{2}(A\cos\theta_K-A'\sin{\theta_K}),\nonumber\\
&&g_{K^*_0(1430)\pi K}=\sqrt{2}(A\sin\theta_K+A'\cos{\theta_K}),\\
&&g_{f_0(M)\pi\pi}=2(-AR_{f_0(M)NN}+\sqrt{2}A'R_{f_0(M)N'}+2A''R_{f_0(M)G}),
\nonumber
\end{eqnarray}
\begin{eqnarray}
&&g_{f_0(M)K\bar{K}}=\sqrt{2}(-AR_{f_0(M)NS}+A'R_{f_0(M)N'}+\sqrt{2}A'R_{f_0(M)S'}
+2\sqrt{2}A''R_{f_0(M)G}),\nonumber\\
&&g_{f_0(M)\eta\eta}=2\left(-AR_{f_0(M)NS}\cos{\theta_P}\sin{\theta_P}+
\frac12AR_{f_0(M)NN}\cos^2{\theta_P}\right.\nonumber\\
&&\hspace{3cm}\left.+\frac{1}{\sqrt{2}}A'R_{f_0(M)N'}\cos^2{\theta_P}+
A'R_{f_0(M)S'}\sin^2{\theta_P}+A''R_{f_0(M)G}\right),\nonumber\\
&&g_{f_0(M)\eta\eta'}=2\left(AR_{f_0(M)NS}\cos{2\theta_P}+
\frac12AR_{f_0(M)NN}\sin{2\theta_P}\right.\nonumber\\
&&\hspace{3cm}\left.+\frac{1}{\sqrt{2}}A'R_{f_0(M)N'}\sin2{\theta_P}-
A'R_{f_0(M)S'}\sin{2\theta_P}\right),\nonumber\\
&&g_{f_0(M)\eta'\eta'}=2\left(AR_{f_0(M)NS}\cos{\theta_P}\sin{\theta_P}+
\frac12AR_{f_0(M)NN}\sin^2{\theta_P}\right.\nonumber\\
&&\hspace{3cm}\left.+\frac{1}{\sqrt{2}}A'R_{f_0(M)N'}\sin^2{\theta_P}+
A'R_{f_0(M)S'}\cos^2{\theta_P}+A''R_{f_0(M)G}\right),\nonumber
\end{eqnarray}
where $\theta_P$ is the $\eta\mbox{-}\eta'$ mixing angle related to the traditional 
octet-singlet mixing angle $\theta_{0\mbox{-}8}$ as $\theta_P=\theta_{0\mbox{-}8}+
54.7^\circ$.
\par
Decay widths for these scalar mesons are expressed by using the coupling constant 
$g_{SPP}$ as 
\begin{eqnarray}
&&\Gamma(a_0(M)\to K^+(m_1)K^-(m_2)+K^0(m_1)\overline{K^0}(m_2))=2
\frac{g^2_{a_0(M)K\bar{K}}}{8\pi}\frac{|{\bi q}|}{m^2_{a_0(M)}}\left[
\left(\frac{m^2_{Mm_1m_2}}{2}\right)^2\right],\nonumber\\
&&\Gamma(a_0(M)\to \pi(m_1)+\eta(m_2))=\frac{g^2_{a_0(M)\pi\eta}}{8\pi}
\frac{|{\bi q}|}{m^2_{a_0(M)}}\left[\left(\frac{m^2_{Mm_1m_2}}{2}\right)^2\right],
\nonumber\\
&&\Gamma(a_0(M)\to \pi(m_1)+\eta'(m_2))=\frac{g^2_{a_0(M)\pi\eta'}}{8\pi}
\frac{|{\bi q}|}{m^2_{a_0(M)}}\left[\left(\frac{m^2_{Mm_1m_2}}{2}\right)^2
\right],\nonumber\\
&&\Gamma({K^*_0}^+(M)\to \pi^+(m_1)K^0(m_2)+\pi^0(m_1)K^+(m_2))=3\frac{g^2_{K^*_0(M)
\pi K}}{8\pi}\frac{|{\bi q}|}{m^2_{K^*_0(M)}}\left[\left(\frac{m^2_{Mm_1m_2}}{2}
\right)^2\right],\nonumber\\
&&\Gamma(f_0(M)\to \pi^+(m_1)\pi^-(m_2)+\pi^0(m_1)\pi^0(m_2))=
\frac32\frac{g^2_{f_0(M)\pi\pi}}{8\pi}\frac{|{\bi q}|}{m^2_{f_0(M)}}
\left[\left(\frac{m^2_{Mm_1m_2}}{2}\right)^2\right],\\
&&\Gamma(f_0(M)\to K^+(m_1)K^-(m_2)+K^0(m_1)\overline{K^0}(m_2))=
2\frac{g^2_{f_0(M)K\bar{K}}}{8\pi}\frac{|{\bi q}|}{m^2_{f_0(M)}}
\left[\left(\frac{m^2_{Mm_1m_2}}{2}\right)^2\right],\nonumber\\
&&\Gamma(f_0(M)\to \eta(m_1)+\eta(m_2))=
2\frac{g^2_{f_0(M)\eta\eta}}{8\pi}\frac{|{\bi q}|}{m^2_{f_0(M)}}
\left[\left(\frac{m^2_{Mm_1m_2}}{2}\right)^2\right],\nonumber\\
&&\Gamma(f_0(M)\to \eta(m_1)+\eta'(m_2))=
\frac{g^2_{f_0(M)\eta\eta'}}{8\pi}\frac{|{\bi q}|}{m^2_{f_0(M)}}
\left[\left(\frac{m^2_{Mm_1m_2}}{2}\right)^2\right],\nonumber\\
&&\Gamma(f_0(M)\to \eta'(m_1)+\eta'(m_2))=
2\frac{g^2_{f_0(M)\eta'\eta'}}{8\pi}\frac{|{\bi q}|}{m^2_{f_0(M)}}
\left[\left(\frac{m^2_{Mm_1m_2}}{2}\right)^2\right].\nonumber
\end{eqnarray}
Here $|{\bi q}|$ and $m_{Mm_1m_2}$ are defined as 
$$ 
\begin{array}{l}
|{\bi q}|=\dis{\sqrt{\left(\frac{M^2+m_2^2-m_1^2}{2M}\right)^2-m_2^2}},\\
m_{Mm_1m_2}=\sqrt{M^2-m_1^2-m_2^2},
\end{array}
$$
and for the case $M\approx m_1+m_2$, we use the next formula for $|{\bi q}|$ 
\cite{harada}, 
\begin{equation}
|{\bi q}|={\rm Re}\frac{1}{\sqrt{2\pi}\Gamma_M}\int^{M+\infty}_{M-\infty}
e^{-\frac{(m-M)^2}{2\Gamma^2_M}}\times\sqrt{\left(\frac{M^2+m_2^2-m_1^2}{2M}
\right)^2-m_2^2}\,dm,
\end{equation} 
where $\Gamma_M$ is the decay width of particle with mass $M$.
\par
We use the experimental data of the scalar meson decays cited in Ref. 
\cite{PDG 2006}; 
$\Gamma(a_0(980)\to \pi\eta+K\bar{K})=75\pm25{\rm MeV}$, 
$\Gamma(a_0(1450)\to \pi\eta+\pi\eta'+K\bar{K})=265\pm13{\rm MeV}$, 
$\Gamma(K^*_0(1430)\to\pi K)=270\pm43{\rm MeV}$ \cite{ASTON}, 
$\Gamma(f_0(980)\to \pi\pi+K\bar{K})=70\pm30{\rm MeV}$, 
$\Gamma(f_0(1370)\to \pi\pi+K\bar{K}+\eta\eta)=214\pm120{\rm MeV}$ \cite{BUGG}, 
$\Gamma(f_0(1500)\to \pi\pi+K\bar{K}+\eta\eta+\eta\eta')=
55\pm9{\rm MeV}$ \cite{AMSLER}, 
$\Gamma(f_0(1710)\to \pi\pi+K\bar{K}+\eta\eta)=
137\pm8{\rm MeV}$, 
for the best fitting of our model parameters, $A$, $A'$, $A''$ and 
$\theta_P$. 
Results are tabulated in Table II for non-derivative coupling case and in Table III 
for derivative coupling case. 
For the $\theta_P$, we search the best fit value in the range $(54.7\pm18)^\circ$. 
We estimate the best fit values of these parameters for various points $(0.2, 0.25, 
0.30, 0.35, 0.40, 0.45){\rm GeV}^2$ of inter-mixing parameter $\lambda_{01}$. 
We show the values of $\Gamma(a_0(980)\to \pi\eta+K\bar{K})$ etc. and $g_{a_(980)\pi
\eta}$ etc. for best fitted $A$, $A'$, $A''$ and $\theta_P$. 
Values in the row just below the one denoting $\Gamma(a_0(980)\to \pi\eta+
K\bar{K})$ etc. denote the experimental values and the values in the row just below 
the one denoting $g_{a_0(980)\pi\eta}$ etc. are the values Eqs.\ (18),\ (18'),\ (27),\ 
(27') obtained from $\phi\to a_0\gamma/\pi^0\eta\gamma$ and $\phi\to f_0\gamma/
\pi^0\pi^0\gamma$ decay analyses. 
\par
These results in Table II (non-derivative coupling case) show that the values of 
$g_{a_0(980)\pi\eta}$ etc. obtained for $\lambda_{01}=0.30\sim0.35($GeV$^2$) are 
close to the values obtained in $\phi\to a_0\gamma/\pi^0\eta\gamma$ and $\phi\to 
f_0\gamma/\pi^0\pi^0\gamma$ decay analyses. 
For the derivative coupling case (showed in Table III), the values of $g_{a_0(980)\pi
\eta}$ etc. except for $g_{f_0(980)K\bar{K}}$ obtained for $\lambda_{01}
=0.30\sim0.35($GeV$^2$) are also close to the values obtained in $\phi\to a_0
\gamma/\pi^0\eta\gamma$ and $\phi\to f_0\gamma/\pi^0\pi^0\gamma$ decay analyses. 
But characteristic feature $g_{f_0(980)K\bar{K}}/g_{a_0(980)K\bar{K}}\sim2$ obtained 
in $\phi\to a_0\gamma/\pi^0\eta\gamma$ and $\phi\to f_0
\gamma/\pi^0\pi^0\gamma$ decay analyses cannot be taken in any values of 
$\lambda_{01}$ for derivative coupling case.. 
\begin{table}
\begin{footnotesize}
\begin{tabular}{|c||c|c|c|c|c|}
\hline
$\lambda_{01}({\rm GeV}^2)$ & $A$ & $A'$ & $A''$ & $\theta_P$ & 
$\Gamma_{a_0(980)\to\pi\eta+K\bar{K}}$\\     
\cline{2-6}
   & GeV & GeV & GeV &degree$(^\circ)$&$0.075\pm0.025$GeV \\
\hline
$0.20$ & $2.8$   & $1.2$   & $-0.25$   & $36.7$  & $0.189$ \\
     \hline
$0.25$ & $2.5$   & $1.2$   & $-0.23$   & $36.7$  & $0.133$ \\
     \hline
$0.30$ & $2.3$   & $1.2$   & $-0.24$   & $36.7$  & $0.097$ \\
     \hline
$0.35$ & $1.9$   & $1.3$   & $-0.24$   & $50.2$  & $0.081$ \\
     \hline
$0.40$ & $1.7$   & $1.3$   & $-0.24$   & $59.2$  & $0.072$ \\
     \hline
$0.45$ & $1.5$   & $1.4$   & $-0.26$   & $72.7$  & $0.068$ \\
     \hline
    &$\Gamma_{a_0(1450)\to{{\pi\eta+\pi\eta'}\atop{+K\bar{K}}}}$ & $\Gamma_{K_0^*
    (1430)\to\pi K}$& $\Gamma_{f_0(980)\to\pi\pi+K\bar{K}}$ & $\Gamma_{f_0(1370)
    \to{{\pi\pi+K\bar{K}}\atop{+\eta\eta}}}$
& $\Gamma_{f_0(1500)\to{{\pi\pi+K\bar{K}}\atop{+\eta\eta+\eta\eta'}}}$  \\
\cline{2-6}
   & $0.265\pm0.013$GeV & $0.270\pm0.043$GeV & $0.070\pm0.030$GeV &$0.214\pm0.120$
   GeV   &$0.055\pm0.009$GeV   \\
\hline  
$0.20$ & $0.242$  & $0.192$  & $0.119$  & $0.034$ & $0.063$ \\
     \hline
$0.25$ & $0.250$  & $0.204$  & $0.107$  & $0.031$ & $0.057$ \\
     \hline
$0.30$ & $0.258$  & $0.214$  & $0.104$  & $0.029$ & $0.055$ \\
     \hline
$0.35$ & $0.273$  & $0.232$  & $0.098$  & $0.034$ & $0.058$ \\
     \hline
$0.40$ & $0.263$  & $0.233$  & $0.098$  & $0.040$ & $0.054$ \\
     \hline
$0.45$ & $0.272$  & $0.253$  & $0.124$  & $0.084$ & $0.056$ \\
     \hline
    &$\Gamma_{f_0(1710)\to{{\pi\pi+K\bar{K}}\atop{+\eta\eta}}}$ & $g_{a_0(980)K
    \bar{K}}$& $g_{a_0(980)\pi\eta}$ & $g_{f_0(980)K\bar{K}}$& $g_{f_0(980)\pi\pi}$\\
\cline{2-6}
   & $0.137\pm0.008$GeV & $2.18\pm0.12$GeV & $1.89\pm0.75$GeV & $4.72\pm0.82$GeV 
   & $1.12\pm0.69$GeV\\
\hline  
$0.20$ & $0.177$  & $3.62$  & $2.84$  & $4.51$  & $0.31$ \\
     \hline
$0.25$ & $0.156$  & $3.09$  & $2.34$  & $4.21$  & $0.45$ \\
     \hline
$0.30$ & $0.140$  & $2.70$  & $1.95$  & $4.05$  & $0.61$ \\
     \hline
$0.35$ & $0.151$  & $2.00$  & $2.06$  & $3.74$  & $0.78$ \\
     \hline
$0.40$ & $0.129$  & $1.59$  & $2.06$  & $3.63$  & $1.00$ \\
     \hline
$0.45$ & $0.141$  & $1.12$  & $2.13$  & $3.66$  & $1.27$ \\
     \hline     
\end{tabular}
\end{footnotesize}
\caption{
The results of the best fit analyses for non-derivative coupling case. 
The experimental data of the scalar meson decay widths used are cited in Ref. 
\cite{PDG 2006}. 
For the $\theta_P$, we search the best fit value in the range $(54.7\pm18)^\circ$. 
The values of $g_{a_(980)\pi\eta}$ etc. are ones obtained for $\phi\to a_0(980)
\gamma/\pi^0\eta\gamma$ and $\phi\to f_0(980)\gamma/\pi^0\pi^0\gamma$ decay analysis. 
}
\end{table}
\begin{table}
\begin{footnotesize}
\begin{tabular}{|c||c|c|c|c|c|}
\hline
$\lambda_{01}({\rm GeV}^2)$ & $A$ & $A'$ & $A''$ & $\theta_P$ & 
$\Gamma_{a_0(980)\to\pi\eta+K\bar{K}}$\\     
\cline{2-6}
   & GeV$^{-1}$ & GeV$^{-1}$ & GeV$^{-1}$ &degree$(^\circ)$&$0.075\pm0.025$GeV \\
\hline
$0.20$ & $5.8$   & $1.2$  & $-0.26$   & $41.2$  & $0.093$ \\
     \hline
$0.25$ & $8.3$   & $0.57$  & $-0.37$   & $36.7$  & $0.170$ \\
     \hline
$0.30$ & $7.2$   & $0.51$  & $-0.33$   & $36.7$  & $0.124$ \\
     \hline
$0.35$ & $6.0$   & $0.54$   & $-0.35$   & $36.7$  & $0.068$ \\
     \hline
$0.40$ & $5.1$   & $0.57$   & $-0.36$   & $36.7$  & $0.054$ \\
     \hline
$0.45$ & $4.2$   & $0.59$   & $-0.34$   & $50.2$  & $0.051$ \\
     \hline
    &$\Gamma_{a_0(1450)\to{{\pi\eta+\pi\eta'}\atop{+K\bar{K}}}}$ & $\Gamma_{K_0^*
    (1430)\to\pi K}$& $\Gamma_{f_0(980)\to\pi\pi+K\bar{K}}$ & $\Gamma_{f_0(1370)
    \to{{\pi\pi+K\bar{K}}\atop{+\eta\eta}}}$
& $\Gamma_{f_0(1500)\to{{\pi\pi+K\bar{K}}\atop{+\eta\eta+\eta\eta'}}}$  \\
\cline{2-6}
   & $0.265\pm0.013$GeV & $0.270\pm0.043$GeV & $0.070\pm0.030$GeV &$0.214\pm0.120$GeV
   &$0.055\pm0.009$GeV   \\
\hline  
$0.20$ & $0.276$  & $0.239$  & $0.025$  & $0.009$ & $0.057$ \\
     \hline
$0.25$ & $0.274$  & $0.262$  & $0.045$  & $0.083$ & $0.068$ \\
     \hline
$0.30$ & $0.279$  & $0.267$  & $0.036$  & $0.099$ & $0.053$ \\
     \hline
$0.35$ & $0.279$  & $0.266$  & $0.028$  & $0.090$ & $0.055$ \\
     \hline
$0.40$ & $0.279$  & $0.266$  & $0.025$  & $0.082$ & $0.059$ \\
     \hline
$0.45$ & $0.276$  & $0.246$  & $0.022$  & $0.056$ & $0.057$ \\
     \hline
    &$\Gamma_{f_0(1710)\to{{\pi\pi+K\bar{K}}\atop{+\eta\eta}}}$ & $g_{a_0(980)K
    \bar{K}}$& $g_{a_0(980)\pi\eta}$ & $g_{f_0(980)K\bar{K}}$
& $g_{f_0(980)\pi\pi}$  \\
\cline{2-6}
   & $0.137\pm0.008$GeV & $9.04\pm0.50$GeV$^{-1}$ & $5.79\pm2.32$GeV$^{-1}$ 
   & $20.0\pm3.48$GeV$^{-1}$    & $2.43\pm1.50$GeV$^{-1}$\\
\hline  
$0.20$ & $0.158$  & $7.80$  & $7.10$  & $8.63$  & $0.39$ \\
     \hline
$0.25$ & $0.106$  & $11.3$  & $9.42$  & $11.6$  & $0.65$ \\
     \hline
$0.30$ & $0.152$  & $9.65$  & $8.02$  & $9.94$  & $0.87$ \\
     \hline
$0.35$ & $0.152$  & $7.85$  & $6.45$  & $8.23$  & $1.03$ \\
     \hline
$0.40$ & $0.152$  & $6.45$  & $5.24$  & $6.94$  & $1.28$ \\
     \hline
$0.45$ & $0.132$  & $5.07$  & $5.45$  & $5.67$  & $1.39$ \\
     \hline     
\end{tabular}
\end{footnotesize}
\caption{
The results of the best fit analyses for derivative coupling case. 
The experimental data of the scalar meson decay widths used are cited in Ref. 
\cite{PDG 2006}. 
For the $\theta_P$, we search the best fit value in the range $(54.7\pm18)^\circ$. 
The values of $g_{a_(980)\pi\eta}$ etc. are ones obtained for $\phi\to a_0(980)
\gamma/\pi^0\eta\gamma$ and $\phi\to f_0(980)\gamma/\pi^0\pi^0\gamma$ decay 
analysis. 
}
\end{table}
\section{Conclusion}
From the invariant mass distribution analysis of radiative decays $\phi\to 
a_0(980)\gamma\to\pi^0\eta\gamma$ and $\phi\to f_0(980)\gamma\to\pi^0\pi^0\gamma$, 
we obtain the results that the $K\bar{K}$ loop diagram contribution for $\phi a_0
\gamma$ and $\phi f_0\gamma$ couplings are dominant for both non-derivative 
and derivative $SPP$ coupling cases. 
We assume that $\phi\to a_0(980)\gamma$ and $\phi\to f_0(980)\gamma$ decays are 
caused through only the $K\bar{K}$ loop diagram, and then we get the results 
Eqs. (18), (18'), (27), (27') of $SPP$ coupling constants, \\
\\
\hspace{3cm}
\begin{tabular}{c|c|c}
               &non-derivative coupling&\ \ \ \ derivative coupling
               \ \ \ \ \\
\hline               
$g_{a K\bar{K}}$&$2.18\pm0.12\ {\rm GeV}$& $9.04\pm0.50\ {\rm GeV^{-1}}$ \\
$g_{a_0\pi\eta}$&$1.89\pm0.75\ {\rm GeV}$& $5.79\pm2.32\ {\rm GeV^{-1}}$ \\
$g_{f K\bar{K}}$&$4.72\pm0.82\ {\rm GeV}$& $20.0\pm0.50\ {\rm GeV^{-1}}$ \\
$g_{f_0\pi\pi}$ &$1.12\pm0.69\ {\rm GeV}$& $2.43\pm1.50\ {\rm GeV^{-1}}$
\end{tabular}\\
\par 
We consider that the scalar $a_0(980)$ and $f_0(980)$ are $qq\bar{q}\bar{q}$ states 
and mix with high mass scalar mesons considered as $q\bar{q}$ states. 
The low mass scalar and high mass scalar mesons are considered to mix through the 
inter-mixing parameter $\lambda_{01}$. 
In our mass formula, we obtain the mixing angle $\theta_a$, $\theta_K$ and mixing 
parameters $R_{f_0(980)NN}$'s using the mass values of low mass scalar mesons 
($a_0(980)$, $K_0^*(800)$, $f_0(600)$, $f_0(980)$) and high mass scalar mesons 
and glueball ($a_0(1450)$, $K_0^*(1430)$, $f_0(1370)$, $f_0(1500)$, $f_0(1710)$).
We tabulate these values for $\lambda_{01}=(0.30\lra0.35)\,$GeV$^2$. 
$$\theta_a=(15.0\lra17.8)^{\circ},\ \ \ \theta_K=(13.8\lra16.4)^\circ\hspace{5cm},
$$
\hspace{0.1cm}
\begin{tabular}{c|ccccc}
           & $f_{NN}$ & $f_{NS}$ & $f_{N'}$ & $f_{S'}$ & $f_G$ \\
\hline
$f_0(600)$ & $-0.98\lra-0.97$ & $0.05\lra0.07$ & $0.20\lra0.23$ & $-0.06\lra-0.08$ 
& $-0.00\lra-0.01$\\
$f_0(980)$ & $-0.05\lra-0.07$ & $-0.93\lra-0.90$ & $0.11\lra0.14$ & $0.34\lra0.40$ 
& $\sim-0.02$\\
$f_0(1370)$ & $0.13\lra0.16$ & $-0.25\lra-0.29$ & $0.48\lra0.49$ 
&$-0.83\lra-0.80$ & $\sim0.02$\\
$f_0(1500)$ & $-0.02\lra-0.03$ & $-0.03\lra-0.05$ & $-0.09\lra-0.10$ 
&$-0.02\lra-0.03$ & $\sim0.99$\\
$f_0(1710)$ &$-0.16\lra-0.19$ & $-0.25\lra-0.30$ & $-0.85\lra-0.82$ 
&$-0.44\lra-0.43$ & $-0.10\lra-0.12$
\end{tabular} \\
\\
The fact that $f_0(980)$ state considered as the $f_{NS}$ state mainly has the 
rather large $f_{S'}$ component with sign opposite to the $f_{NS}$ one suggests a 
possibility that $g_{f_0K\bar{K}}/g_{a_0K\bar{K}}$ can be about 2, because 
$g_{f_0K\bar{K}}/g_{a_0K\bar{K}}=|(-AR_{f_0NS}+A'R_{f_0N'}+\sqrt{2}A'R_{f_0S'}+
A''R_{f_0G})/(A\cos{\theta_a}-A'\sin{\theta_a})|$ and $\theta_a>0$. 
In our model, $f_0(1500)$ meson is considered as glueball, and $f_0(1370)$ meson is 
almost $f_{S'}$ state with rather large $f_{N'}$ component and $f_0(1710)$ meson is 
almost $f_{N'}$ state with rather large $f_{S'}$ component. 
\par
Because $g_{f_0K\bar{K}}$'s are related to the mixing parameters $R_{f_0NS}$'s 
and coupling strengths $A$, $A'$, $A''$ and $\eta\mbox{-}\eta'$ mixing angle 
$\theta_P$, we executed the best fit analyses of the various $SPP$ decays in 
the wide range of the $\lambda_{12}$ value for non-derivative and 
derivative coupling cases. 
The best fit values of $A$'s and $g_{a_0K\bar{K}}$'s are tabulated in Table II 
and III. 
These results suggest that the non-derivative coupling seems to be reasonable than 
the derivative one and the inter-mixing parameter $\lambda_{12}$ is rather large 
$0.30\lra0.35$.    

\end{document}